\RequirePackage{fix-cm}

\documentclass{article}

\usepackage{graphicx}
\usepackage{stmaryrd}
\usepackage{amssymb}
\usepackage{amsmath}
\usepackage{eucal}

\usepackage[ruled,vlined]{algorithm2e}
\usepackage[driverfallback=dvipdfm,breaklinks=true,pdfcreator={LaTeX e Dvips}]{hyperref}
\usepackage{float}

\usepackage{mathptmx}
\usepackage{natbib}
\usepackage{color}
\usepackage{ulem}
\usepackage{booktabs}
\usepackage{multirow}
\usepackage{textcomp}

\newcommand{\cl}{\color}

\title{Topology design of two-fluid heat exchange}

\author{Hiroki Kobayashi$^\text{a}$,
        Kentaro Yaji$^\text{a,}\footnote{Corresponding author: {\tt yaji@mech.eng.osaka-u.ac.jp} (Kentaro Yaji)}$ ,
        Shintaro Yamasaki$^\text{a}$,
        Kikuo Fujita$^\text{a}$ \\[12pt]
$^\text{a}$\textit{Department of Mechanical Engineering, Graduate School
              of Engineering,}\\
              \textit{Osaka University, 2-1, Yamadaoka,
              Suita, Osaka 565-0871, Japan}}

\begin{document}

\maketitle

\begin{abstract}
Heat exchangers are devices that typically transfer heat between two fluids. 
 The performance of a heat exchanger such as heat transfer rate and pressure loss strongly depends on the flow regime in the heat transfer system.
In this paper, we present a density-based topology optimization method for a two-fluid heat exchange system, which achieves a maximum heat transfer rate under fixed pressure loss.
We propose a representation model accounting for three states, i.e., two fluids and a solid wall between the two fluids, by using a single design variable field.
The key aspect of the proposed model is that mixing of the two fluids can be essentially prevented without any penalty scheme.
This is because the solid constantly exists between the two fluids due to the use of the single design variable field.
We demonstrate the effectiveness of the proposed approach through three-dimensional numerical examples in which an optimized design is compared with a simple reference design, and the effects of design conditions (i.e., Reynolds number, Prandtl number, design domain size, and flow arrangements) are investigated.
\flushleft
\textbf{Keywords}\ \ Topology optimization $\cdot$ Two kinds of fluids $\cdot$ Heat exchange $\cdot$ Interpolation scheme
\end{abstract}

\section{Introduction}
\label{sec_intro}
Heat exchangers are devices that transfer heat between two or more fluids. Recently, designing high-performance heat exchangers has been prioritized owing to an increased demand for such systems with low energy consumption. Most heat exchangers involve two-fluid heat exchange with indirect contact, i.e., there are two fluids separated by a wall. 
In such heat exchangers, the performances depend on the flow regime. 
For instance, the curving and branching of the flow channels of each fluid and the adjacency of the flow channels to each other have a significant effect on the heat transfer rate and the pressure loss. 
Therefore, consideration of such characteristics is significant in heat exchanger design. 

Generally, in heat exchanger design, a type is first selected from the existing ones, and the details, e.g., size, shape, and fin type, are determined \citep{shah2003fundamentals}. 
Optimization methods for the existing types of heat exchangers have been well researched in previous works. 
For instance, \cite{hilbert2006multi} studied a multiobjective optimization method concerning the blade shape of a tubular heat exchanger via a genetic algorithm.  
\cite{kanaris2009optimal} applied a response surface method to optimization problems for a plate heat exchanger with undulated surfaces. 
 \cite{guo2018design} proposed a design method to simultaneously determine the fin types and their optimal sizes in plate fin heat exchangers. 

The existing types of heat exchangers are typically manufactured through traditional processes such as pressing of plates and bending of tubes.
This implies that the typical heat exchangers are designed under strict constraints pertaining to their manufacturability.
However, there is an increasing demand to produce heat exchangers more efficient and compact, as the traditional manufacturing techniques are usually too inefficient to satisfy this demand.
To produce innovative heat exchangers, these constraints pertaining to their manufacturability should be eliminated. 
In this regard, using additive manufacturing is a promising option, and has been employed for manufacturing innovative heat exchangers \citep[e.g.][]{scheithauer2018additive} in instances where traditional techniques cannot be easily be applied. 
According to such progression of additive manufacturing technologies, revealing and understanding the ideal flow regime of the two fluids in heat exchange systems become increasingly important.
However, the ideal flow regime is not obvious in most cases. 
For such a problem, topology optimization is a promising methodology, as it enables shape and topological changes of the structure. 

Topology optimization was first proposed by \cite{bendsoe1988generating}. It essentially comprises introduction of a fixed design domain and replacement of the original optimization problem with a material distribution problem. 
A characteristic function is introduced to represent either solid material or void at the appropriate points in the design domain. 
Although topology optimization was originally applied to solid mechanics problems, it has been developed for application in various physical problems \citep{bendsoe2003topology,sigmund2013topology,deaton2014survey}.

For fluid flow problems, \cite{borrvall2003topology} proposed a topology optimization method for solid/fluid distribution problems in Stokes flow. 
The critical feature in their method is the introduction of a fictitious body force term, which can discriminate fluid and solid at points in the design domain, according to each design variable value. 
Their method was extended for Navier--Stokes equations \citep{gersborg2005topology,olesen2006high}, and design problems of fluidic devices, such as microreactors \citep{okkels2007scaling}, micromixers \citep{andreasen2009topology,deng2018topology}, Tesla valves \citep{lin2015topology,sato2017topology}, and flow batteries \citep{yaji2018topology,chen2019computational}. 
The thermal-fluid problems that have been studied include forced convection problems \citep{matsumori2013topology,koga2013development,yaji2015topology}, natural convection problems \citep{alexandersen2014topology,coffin2016level}, and large-scale three-dimensional problems involving cluster computing \citep{alexandersen2016large,yaji2018large}.
In addition, \cite{dilgen2018density} applied density-based topology optimization to turbulent heat transfer problems. 
For practical approaches in utilizing topology optimization in complex fluid problems, simplified models have recently been studied for solving turbulent heat transfer problems \citep{zhao2018poor,yaji2020multifidelity,kambampati2020level}, natural convection problems \citep{asmussen2019poor,pollini2020poor}, and heat exchanger design problems  \citep{haertel2017fully,kobayashi2019freeform}. 
The history and the state-of-the-art of topology optimization for fluid problems including heat transfer problems are summarized in the recently published review paper \citep{alexandersen2020review}. 

Many studies regarding topology optimization of thermal-fluid problems dealt with solid/fluid distribution, i.e., only one fluid is considered. 
A topology optimization method for two-fluid and one-solid distribution problems is essential to generate an innovative design of a two-fluid heat exchanger. 
\cite{tawk2019topology} has recently proposed a topology optimization method for heat transfer problems involving two-fluid and one-solid domains.
In their method, two types of design variable fields are introduced. The first design variable field expresses either solid or fluid, and the second one expresses the ratio of the two fluids. The interpolation function in \cite{borrvall2003topology} is extended for two-fluid problems using an interpolation function for multi-material topology optimization presented in the literature \citep{bendsoe1999material}. 
In addition, a penalty function is introduced to prevent the mixing of the two fluids, as an essential constraint. 
The proposed approach was applied to two-dimensional examples, and it was revealed that the optimized designs realized an increasing heat transfer rate and a decreasing pressure loss.
The authors \citep{tawk2019topology}, however, stated that the solution search becomes unstable in certain conditions because of the penalty scheme. 

In the general framework of multi-material topology optimization, three states (two types of materials and void) are accounted for by using multiple design variable fields. 
Conversely, in topology optimization of two-fluid heat exchange, the two fluids should be separated by a wall.
For two-fluid problems, solving the multi-material topology optimization typically requires a penalty scheme as with the previous work \citep{tawk2019topology}, since the representation model for the multiple states has an excessive degree of freedom. 
There arises a possibility that a simpler representation model can be obtained by focusing on the characteristics of the two-fluid problems.
For instance, \cite{detopology} proposed a representation model for design-dependent pressure load problems, in which three states (interior and exterior fluids, and a wall) are represented using a single design variable field.  
They demonstrated that the leakage between the two fluids can be circumvented by only using the single design variable field, whose middle value corresponds to the wall between the fluids.

In this study, we propose a representation model for topology optimization of two-fluid heat exchange. 
The novel feature of the proposed model is the representation of two different types of fluids and a wall by using a single design variable field, whereas, in the previous work \citep{detopology}, the two fluids used were required to be of the same type. 
Therefore, we introduce two types of fictitious force terms in each flow field for representing a wall between the two types of fluid, which essentially cannot mix.
The topology optimization problem is formulated as a maximization problem of the heat transfer rate under a fixed pressure loss.
We emphasize that the two-fluid heat exchange problems should be treated as three-dimensional problems since the topological change is significantly restricted in the case of two-dimensional problems. 
Thus, we apply the proposed approach to three-dimensional heat exchange problems and examine the effects of several design conditions, such as Reynolds number, Prandtl number, design domain size, and flow arrangements. 

The remaining sections of this paper are organized as follows. In Section~\ref{sec_applicability}, the design problem of two-fluid heat exchange is presented, as well as the method to represent two fluids and one solid by the single design variable field is briefly discussed. 
In Section~\ref{sec_to}, the analysis model and the proposed representation model is presented, and the topology optimization problem is formulated. Furthermore, the numerical implementation is presented. In Section~\ref{sec_numerical}, several three-dimensional numerical examples are provided to validate the proposed method and to discuss the ideal flow regime of the two-fluid heat exchange. Finally, Section~\ref{sec_conclusion} concludes this study. 

\section{Topology optimization for two-fluid heat exchange\label{sec_applicability}}

\subsection{Design problem\label{ssec_dp}}
\begin{figure}[tbp]
\centering
\includegraphics[width=80mm]{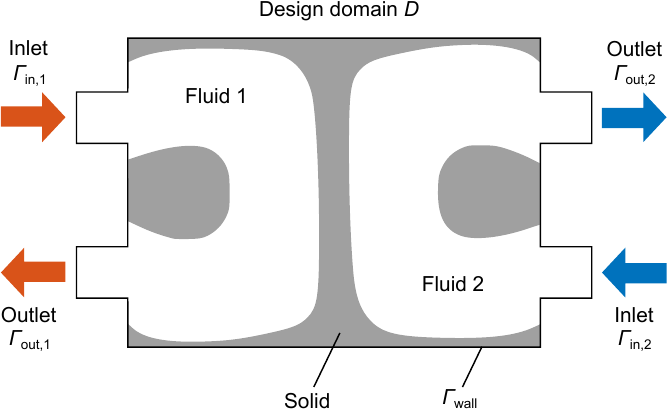}
\caption{Schematic diagram of two-fluid heat exchange\label{fig1}}
\end{figure}

A simple heat exchanger, in which flow channels of the two fluids are separated by a wall as presented in Fig.~\ref{fig1}, is considered. An inlet and an outlet of each fluid are respectively set, and each fluid flows independently. The heat is transferred between the two fluids in the design domain $D$. The hot and cold fluids are denoted by ``fluid 1'' and ``fluid 2,'' respectively.

In this study, we consider a topology optimization problem for designing the shape and topology of the flow channels and the separating wall in the design domain $D$. 
The objectives of the heat exchanger design are the maximization of the heat transfer rate and minimization of the pressure loss. 
This multi-objective design problem is transformed into a single-objective design problem by fixing the pressure loss. 
The detailed formulation is provided in Section \ref{sec_to}. 

\subsection{Basic concept of topology optimization}
Topology optimization essentially involves the replacement of the original structural optimization problem with a material distribution problem in a fixed design domain $D$ that contains the original design domain $\Omega$. \cite{bendsoe1988generating} introduced a characteristic function $\chi_{\Omega}$ that assumes a discrete value 0 and 1, as follows:
\begin{equation}
\chi_{\Omega} (\mathbf{x}) =
\begin{cases}
1\quad \mathrm{if}\quad \mathbf{x} \in  \Omega\\
0\quad \mathrm{if}\quad \mathbf{x} \in  D\backslash \Omega,
\end{cases}
\end{equation}
where $\mathbf{x}$ represents the position in the fixed design domain $D$. Since the characteristic function $\chi_{\Omega}$ is a discontinuous function, a relaxation technique is required for gradient-based optimization. 
In the density-based method \citep{bendsoe2003topology}, the characteristic function $\chi_{\Omega}$ is replaced with a continuous scalar function $0 \leq \psi(\mathbf{x}) \leq 1$, which corresponds to the raw design variable field in the topology optimization problem. 

To ensure the spatial smoothness of the material distribution in $D$, the filtering technique is generally applied in topology optimization. 
In this study, Helmholtz-type filter \citep{kawamoto2011heaviside,lazarov2011filters} is employed, as follows: 
\begin{equation}
-R^2 \nabla^2 \gamma +\gamma = \psi ,
\label{eqn_filter}
\end{equation}
where $\gamma$ is a smoothed design variable field, and $R$ is a filtering radius. 
By solving (\ref{eqn_filter}), $\psi$ is mapped to $\gamma$, thus ensuring a spatial smoothness in $D$. 

\begin{table*}[tbp]
  \centering
  \caption{Representation model of two fluids and one solid based on the multi-material topology optimization \citep{tawk2019topology}\label{table1}}
    \begin{tabular}{cc|ccc}
    \toprule
          &       & \multicolumn{3}{c}{Design variable 1} \\
          &       & $\gamma_1=0$ & $0<\gamma_1 <1$ & $\gamma_1=1$ \\
    \midrule
    \multicolumn{1}{c}{\multirow{3}[2]{*}{{\begin{tabular}{c} Design \\ variable 2 \end{tabular}}}} & $\gamma_2=0$ & Solid & Mixture of solid and fluid 1 & Fluid 1 \\
          & $0 < \gamma_2 <1$ & Solid &    \begin{tabular}{c} Mixture of solid, fluid 1, and fluid 2 \end{tabular}   & Mixture of fluid 1 and fluid 2 \\
          & $\gamma_2=1$ & Solid & Mixture of solid and fluid 2 & Fluid 2 \\
    \bottomrule
    \end{tabular}%
\end{table*}%

\begin{table}[tbp]
  \centering
  \caption{Representation model of two fluids and one solid in the proposed method\label{table2}}
    \begin{tabular}{ccc}
    \toprule
    \multicolumn{3}{c}{Design variable} \\
    $\gamma=0$ & $0<\gamma <1$ & $\gamma=1$ \\
    \midrule
    Fluid 2 & Solid & Fluid 1 \\
    \bottomrule
    \end{tabular}%
\end{table}%

\subsection{Representation model}
For the topology optimization of two-fluid heat exchange, the design variable field should represent the two fluids and the one solid, as shown in Fig.~\ref{fig1}. 
Here, we discuss the method to represent the two fluids and the one solid by the design variable field. 

In the previous work regarding topology optimization of two fluids \citep{tawk2019topology}, the method presented was based on multi-material topology optimization. The representation of the two fluids by design variable fields are presented in Table~\ref{table1}. 
Here, design variable 1, corresponding to $\gamma_1$, represents the ratio of the solid and the fluids, 
whereas design variable 2, corresponding to $\gamma_2$, represents the ratio of fluid 1 and 2. 

Multi-material topology optimization is extensively studied regarding distribution problems of materials and void \citep{bendsoe1999material}. In the typical multi-material topology optimization, two types of design variable fields are utilized to represent two types of materials and void. In this regard, the distribution problem of two fluids and one solid is similar in that the three states are accounted for. However, it differs in the adjacency of the three states. In the two-fluid heat exchange problems, fluids 1 and 2 should not be adjacent and mixed. As revealed in Table~\ref{table1}, fluid 1 and 2 can be adjacent and mixed when $\gamma_1=1 $ and $ 0 < \gamma_2 < 1$. Since the penalty scheme is required to prevent mixing and adjoining of the two fluids, the method based on the multi-material topology optimization is complex and may lead to an unstable solution search. 

The state representation should be simple, and the penalty scheme should not be required for a stable solution search. In this study, we propose a new method that uses a single design variable field. 
Table~\ref{table2} demonstrates the state represented by the design variable field. 
The design variable field $\gamma$ assumes 1 in fluid 1 and 0 in fluid 2. The significant difference to the previous work \citep{tawk2019topology} is to represent the solid in the intermediate value of $\gamma$. This is because the two fluids are separated by the wall in the heat exchange problem. 
The two fluids can consistently be separated by the wall when the spatial smoothness of the design variable field is preserved in the design domain $D$.
It should be noted that a filtering technique (\ref{eqn_filter}), is essential in the proposed representation model. 
In this setting, the mixing of fluids 1 and 2 can be eliminated without using the penalty function. Thus, the main feature of this method is that the two fluids and the one solid can be represented simply by the single design variable field without any penalty schemes.  

\section{Formulation\label{sec_to}}
\subsection{Governing equations and boundary conditions\label{ssec_ge_bc}}
The governing equations of an incompressible steady flow are given as follows:
\begin{equation}
\nabla \cdot \mathbf{u} = 0,
\label{eqn_continuity}
\end{equation}
\begin{equation}
\rho (\mathbf{u} \cdot \nabla) \mathbf{u} = - \nabla p + \mu \nabla^2 \mathbf{u} + \mathbf{F},
\label{eqn_ns}
\end{equation}
where $\mathbf{u}$ is the velocity, $\rho$ is the density, $p$ is the pressure, $\mu$ is the viscosity, and $\mathbf{F}$ is a body force. In the fluid topology optimization, $\mathbf{F}$ is a fictitious body force caused by the solid objects in the flow \citep{borrvall2003topology}. 
The energy equation without the heat source is defined as follows:
\begin{equation}
\rho C_p (\mathbf{u} \cdot \nabla) T - k \nabla^2 T = 0,
\label{eqn_energy}
\end{equation}
where $C_p$ is the specific heat at constant pressure, $T$ is the temperature, and $k$ is the thermal conductivity. 

The boundary conditions are defined as follows:
\begin{align}
p=p_\mathrm{in, 1}, \quad T=T_\mathrm{in,1} \quad \mathrm{on}& \quad \Gamma_{\mathrm{in},1},\label{eqn_bc_1}\\
p=p_\mathrm{out, 1}, \quad \nabla T \cdot \mathbf{n} = 0 \quad \mathrm{on}& \quad \Gamma_{\mathrm{out},1},\label{eqn_bc_2}\\
p=p_\mathrm{in, 2}, \quad T=T_\mathrm{in,2} \quad \mathrm{on}& \quad \Gamma_{\mathrm{in},2},\label{eqn_bc_3}\\
p=p_\mathrm{out, 2}, \quad \nabla T \cdot \mathbf{n} = 0 \quad \mathrm{on}& \quad \Gamma_{\mathrm{out},2},\label{eqn_bc_4}\\
\mathbf{u}=\mathbf{0}, \quad \nabla T \cdot \mathbf{n} = 0 \quad \mathrm{on}& \quad \Gamma_\mathrm{wall}, \label{eqn_bc_5}
\end{align}
where subscripts 1 and 2 represent the fluid 1 and fluid 2, respectively. As mentioned in Section~\ref{ssec_dp}, the constant pressure loss is assumed by fixing the pressure at the inlet and outlet. 

\subsection{Proposed interpolation scheme\label{ssec_interpolation}}
\begin{figure}[tbp]
\centering
\includegraphics[width=80mm]{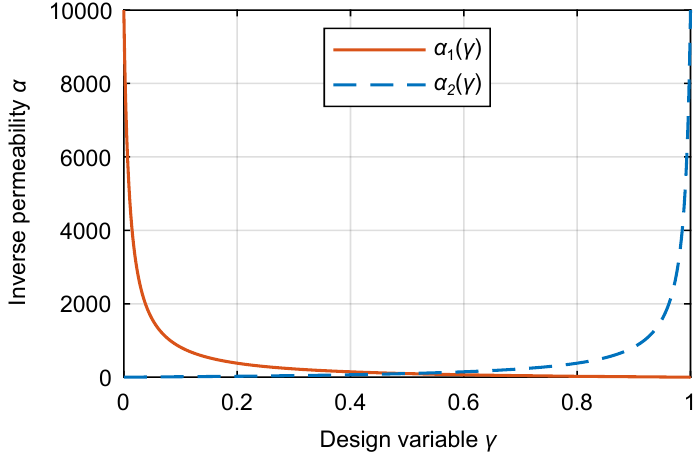}
\caption{Interpolation functions for fluid 1 and 2 at $\alpha_\mathrm{max}= 10^{4}$ and $q=0.01$\label{fig2}}
\end{figure}

In this study, the different fictitious forces in the governing equations for fluid 1 and fluid 2 are imposed separately to realize the state represented by the design variable field indicated in Table~\ref{table2}. Therefore, the governing equations are separated for fluid 1 and fluid 2, as follows:
\begin{equation}
\nabla \cdot \mathbf{u}_1 = 0,
\label{eqn_continuity_summary1}
\end{equation}
\begin{equation}
\rho_1 (\mathbf{u}_1 \cdot \nabla) \mathbf{u}_1 = - \nabla p_1 + \mu_1 \nabla^2 \mathbf{u}_1 + \mathbf{F}_1,
\label{eqn_ns_summary1}
\end{equation}
\begin{equation}
\nabla \cdot \mathbf{u}_2 = 0,
\label{eqn_continuity_summary2}
\end{equation}
\begin{equation}
\rho_2 (\mathbf{u}_2 \cdot \nabla) \mathbf{u}_2 = - \nabla p_2 + \mu_2 \nabla^2 \mathbf{u}_2 + \mathbf{F}_2,
\label{eqn_ns_summary2}
\end{equation}
where subscripts 1 and 2 represent fluid 1 and fluid 2, respectively. The last terms in (\ref{eqn_ns_summary1}) and (\ref{eqn_ns_summary2}) are crucial in the proposed method. These fictitious body forces are defined as follows: 
\begin{equation}
\mathbf{F}_1 = - \alpha_1 \mathbf{u}_1,
\label{eqn_f1}
\end{equation}
\begin{equation}
\mathbf{F}_2 = - \alpha_2 \mathbf{u}_2,
\label{eqn_f2}
\end{equation}
where $ \alpha_1 (\mathbf{x})$ and $ \alpha_2 (\mathbf{x})$ are the inverse permeability of the porous media at position $\mathbf{x}$ \citep{borrvall2003topology}. The fluid hardly flows when $\alpha_1$ or $\alpha_2$ assumes a large value. Consequently, the fluid can flow freely when $\alpha_1$ or $\alpha_2$ assumes a zero value. The inverse permeabilities $\alpha_1$ and $\alpha_2$ are defined as follows:
\begin{equation}
\alpha_1 = \alpha_{\mathrm{max}} \frac{q(1-\gamma)}{q + \gamma},
\label{eqn_alpha1}
\end{equation}
\begin{equation}
\alpha_2 = \alpha_{\mathrm{max}} \frac{q \gamma}{q + 1 - \gamma},
\label{eqn_alpha2}
\end{equation}
where $\alpha_\mathrm{max}$ is the maximum inverse permeability and $q$ is the parameter that controls the convexity of $\alpha$. Both inverse permeabilities are defined based on the interpolation function by \cite{borrvall2003topology}. However, these two interpolation functions are symmetrical with respect to $\gamma$. Figure~\ref{fig2} displays the graph of the interpolation functions at $\alpha_\mathrm{max}= 10^{4}$ and $q=0.01$. As revealed in Fig.~\ref{fig2}, if $\gamma=0$, $\alpha_1$ assumes a large value $\alpha_\mathrm{max}$ and $\alpha_2$ assumes a zero value. For this reason, only fluid 2 can flow whereas fluid 1 cannot flow. Conversely, if $\gamma=1$, only fluid 1 can flow since $\alpha_1$ assumes a zero value and $\alpha_2$ assumes $\alpha_\mathrm{max}$. 
In the intermediate value of $\gamma$, both fluids hardly flow because both $\alpha_1$ and $\alpha_2$ are not zero.
Therefore, the flow characteristics of Table~\ref{table2} can be achieved via these interpolation functions of the inverse permeability. 

To simplify the numerical implementation, the governing equations are replaced with dimensionless form, and the dimensionless quantities, i.e., the Reynolds number and the P\'{e}clet number, are introduced. 

The dimensionless equations of (\ref{eqn_continuity_summary1})--(\ref{eqn_ns_summary2}) are defined as follows:
\begin{equation}
\nabla^* \cdot \mathbf{u}_1^* = 0,
\label{eqn_continuity_dl1}
\end{equation}
\begin{equation}
(\mathbf{u}_1^* \cdot \nabla^*) \mathbf{u}_1^* = - \nabla^* p_1^* + \frac{1}{Re_1} \nabla^{*2} \mathbf{u}_1^* - \alpha_1^* \mathbf{u}_1^*,
\label{eqn_ns_dl1}
\end{equation}
\begin{equation}
\nabla^* \cdot \mathbf{u}_2^* = 0,
\label{eqn_continuity_dl2}
\end{equation}
\begin{equation}
(\mathbf{u}_2^* \cdot \nabla^*) \mathbf{u}_2^* = - \nabla^* p_2^* + \frac{1}{Re_2} \nabla^{*2} \mathbf{u}_2^* - \alpha_2^* \mathbf{u}_2^*,
\label{eqn_ns_dl2}
\end{equation}
where the asterisk symbol represents dimensionless variables defined by a characteristic length $L$, characteristic speeds $U_1$ and $U_2$ as follows:
\begin{align}
\begin{aligned}
\nabla^* = {L} \nabla,\quad {\mathbf{u}}_1^* = \frac{{\mathbf{u}_1}}{{U}_1},\quad {p}_1^* = \frac{{p}_1 - p_\mathrm{out, 1}}{{\rho}_1 {U}_1^2},\\
{\alpha}_1^* = \frac{{L} {\alpha}_1}{{\rho}_1 {U}_1},\quad
\mathit{{Re}}_1 = \frac{{\rho}_1 {U}_1 {L}}{{\mu}_1},
\end{aligned}
\label{eqn_dimensionless_fluid1}
\end{align}
\begin{align}
\begin{aligned}
\nabla^* = {L} \nabla,\quad {\mathbf{u}}_2^* = \frac{{\mathbf{u}_2}}{{U}_2},\quad {p}_2^* = \frac{{p}_2 - p_\mathrm{out, 2}}{{\rho}_2 {U}_2^2},\\
{\alpha}_2^* = \frac{{L} {\alpha}_2}{{\rho}_2 {U}_2},\quad
\mathit{{Re}}_2 = \frac{{\rho}_2 {U}_2 {L}}{{\mu}_2}.
\end{aligned}
\label{eqn_dimensionless_fluid2}
\end{align}
The characteristic speeds $U_1$ and $U_2$ cannot be defined using the magnitude of the inlet velocity because the inlet pressure is fixed, as mentioned in Section~\ref{ssec_ge_bc}. 
Therefore, $U_1$ and $U_2$ are defined using the pressure loss \citep{yaji2015topology}, as follows:
\begin{equation}
{U}_1=\sqrt{\frac{{p}_\mathrm{in, 1}-{p}_\mathrm{out, 1}}{{\rho}_1}},
\label{eqn_U1}
\end{equation}
\begin{equation}
{U}_2=\sqrt{\frac{{p}_\mathrm{in, 2}-{p}_\mathrm{out, 2}}{{\rho}_2}}.
\label{eqn_U2}
\end{equation}
As a result, the proposed interpolation functions $\alpha_1^*(\gamma)$ and $\alpha_2^*(\gamma)$ in the dimensionless form are given as:
\begin{equation}
\alpha_1^* (\gamma) = \alpha_{\mathrm{max}}^* \frac{q(1-\gamma)}{q + \gamma},
\label{eqn_alpha1_dl}
\end{equation}
\begin{equation}
\alpha_2^* (\gamma) = \alpha_{\mathrm{max}}^* \frac{q \gamma}{q + 1 - \gamma},
\label{eqn_alpha2_dl}
\end{equation}
where $\alpha_{\mathrm{max}}^*$ is the maximum inverse permeability.

\begin{figure}[tbp]
\centering
\includegraphics[width=80mm]{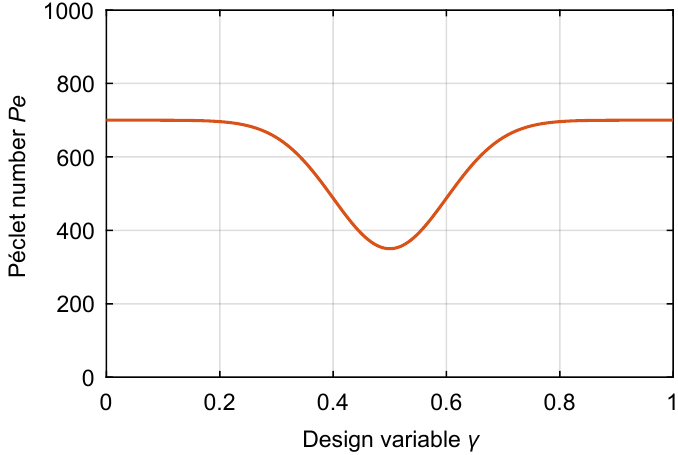}
\caption{Interpolation function for P\'{e}clet number at $Pe_\mathrm{f1}=700$, $Pe_\mathrm{f2}=700$, and $Pe_\mathrm{s}=350$\label{fig3}}
\end{figure}

The dimensionless equation of (\ref{eqn_energy}) is defined as follows:
\begin{equation}
\mathbf{u}^* \cdot \nabla^* T^* - \frac{1}{Pe} \nabla^{*2} T^* = 0.
\label{eqn_energy_dl}
\end{equation}
Here, $Pe$ is the P\'{e}clet number $Pe=Pr Re$, in which $Pr$ is the Prandtl number. 
The dimensionless temperature $T^*$ and the Prandtl number $Pr$ are respectively defined as:
\begin{equation}
{T}^* = \frac{{T} - {T}_\mathrm{in, 2}}{{T}_\mathrm{in, 1} - {T}_\mathrm{in, 2}},\quad \mathit{{Pr}} = \frac{\mu C_p}{k}.
\label{eqn_dimensionless_temp}
\end{equation}
In (\ref{eqn_energy_dl}), since the dimensional parameters of thermal-fluid properties are integrated into the dimensionless parameter, $Pe$, an interpolation function should be introduced so that $Pe$ depends on the states, i.e., fluid 1, fluid 2 or the solid.
Therefore, we propose an interpolation function based on the Gaussian function, as follows:
\begin{align}
\begin{aligned}
Pe(\gamma) = &\left( Pe_\mathrm{s}-\frac{Pe_\mathrm{f1}+Pe_\mathrm{f2}}{2}\right) \exp \left(-\frac{(\gamma-0.5)^2}{2s^2} \right) \\ &+Pe_\mathrm{f2}+(Pe_\mathrm{f1}-Pe_\mathrm{f2})\gamma ,
\label{eqn_gauss_pe}
\end{aligned}
\end{align}
where $s$ is the parameter that controls the shape of the Gaussian function.
$Pe_\mathrm{f1}$, $Pe_\mathrm{f2}$, and $Pe_\mathrm{s}$ are the P\'{e}clet number of fluid 1, fluid 2, and the solid, respectively. As shown in Fig.~\ref{fig3}, $Pr$ in the intermediate region corresponding to the solid can be smaller than that of the two fluids. 
This indicates that the solid has larger thermal conductivity than that of the two fluids, under the condition that the remaining parameters of thermal-fluid properties are fixed.

\subsection{Optimization formulation}
As discussed in Section~\ref{ssec_dp}, the objective of the design problem is to maximize the total heat transfer with a fixed pressure loss. 
The total heat transfer is given as follows:
\begin{equation}
Q=\dot{m} C_p (\overline{T}_{\mathrm{in}} - \overline{T}_{\mathrm{out}}),
\label{eqn_total_heat_exchange}
\end{equation}
where $\dot{m}$ is the mass flow, and $\overline{T}$ is the mean temperature. 
Introducing the boundary integral form, the equation for total heat transfer can be rewritten as:
\begin{equation}
Q= C_p \int_{\Gamma_\mathrm{out}} \rho \mathbf{u} \cdot \mathbf{n} d\Gamma \left(\frac{\int_{\Gamma_\mathrm{in}} (\mathbf{u}\cdot \mathbf{n}) T d\Gamma}{\int_{\Gamma_\mathrm{in}} \mathbf{u}\cdot \mathbf{n} d\Gamma} - \frac{\int_{\Gamma_\mathrm{out}} (\mathbf{u}\cdot \mathbf{n}) T d\Gamma}{\int_{\Gamma_\mathrm{out}} \mathbf{u}\cdot \mathbf{n} d\Gamma} \right).
\end{equation}
Since incompressible flow is assumed in this study, the density is constant and the volume flows at the inlet and outlet are equal. 
In addition, the inlet temperature $T$ is constant as defined in (\ref{eqn_bc_1}) and (\ref{eqn_bc_3}). Thus, the equation can be further rewritten as:
\begin{equation}
Q= \rho C_p \int_{\Gamma_\mathrm{out}} (\mathbf{u}\cdot \mathbf{n}) (T_\mathrm{in} -T) d\Gamma.
\end{equation}
To replace the objective function with dimensionless form, a total heat transfer $Q$ is nondimensionalized as follows:
\begin{equation}
Q^* = \frac{Q}{\rho C_p U L^2 (T_\mathrm{in, 1} - T_\mathrm{in, 2})}.
\label{eq:Q*}
\end{equation}
Based on (\ref{eq:Q*}), we define the objective function in this study as the sum of the total heat transfer in fluid 1 and fluid 2, as follows:
\begin{equation}
J = \int_{\Gamma_{\mathrm{out},1}} (\mathbf{u}^* \cdot \mathbf{n}^*) (1-T^*) d\Gamma  + \int_{\Gamma_{\mathrm{out},2}} (\mathbf{u}^* \cdot \mathbf{n}^*) T^* d\Gamma.
\end{equation}

\begin{figure}[tbp]
\centering
\includegraphics[width=80mm]{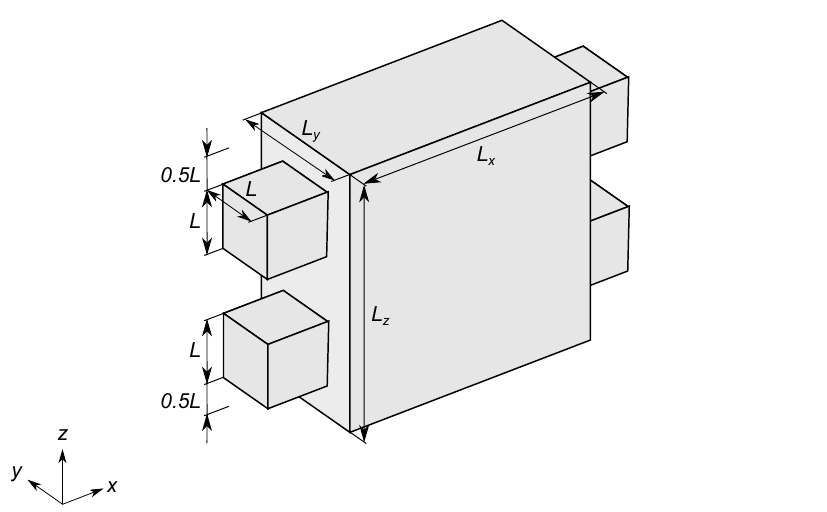}
\caption{Analysis domain and its dimensions for two-fluid heat exchange\label{fig4}}
\end{figure}
\begin{figure}[tbp]
	\centering
	\includegraphics[width=80mm]{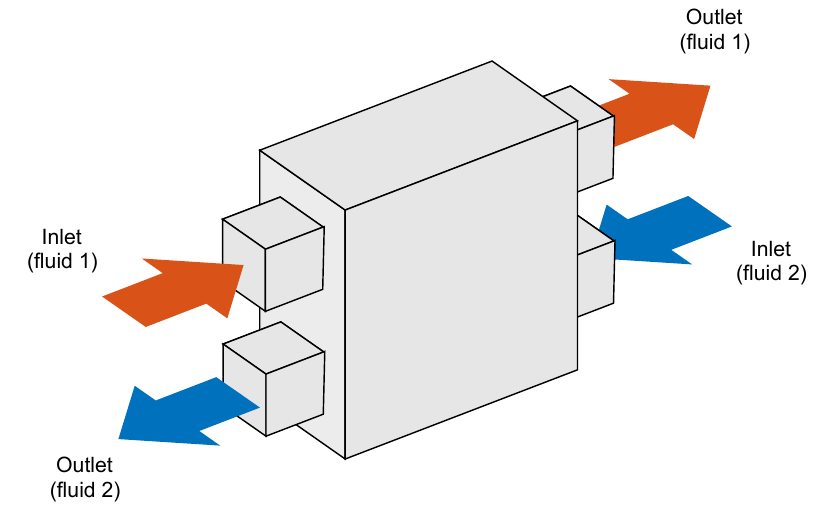}
	\caption{Schematic diagram of counter flow\label{fig5}}
\end{figure}

For brevity, the asterisk symbol of the dimensionless variables is dropped henceforth.

Finally, the topology optimization problem is formulated as follows:
\begin{align}
\begin{aligned}
& \underset{\psi({\mathbf{x}})}{\mathrm{maximize}}\quad {J} = \int_{\Gamma_{\mathrm{out},1}} (\mathbf{u} \cdot \mathbf{n}) (1-T) d\Gamma  + \int_{\Gamma_{\mathrm{out},2}} (\mathbf{u} \cdot \mathbf{n}) T d\Gamma, \\
& \mathrm{subject\ to}\quad 0 \leq \psi({\mathbf{x}}) \leq 1\quad \mathrm{for} \ \forall {\mathbf{x}} \in D,
\label{eqn_optimization_problem}
\end{aligned}
\end{align}
where the thermal-fluid field is governed by (\ref{eqn_continuity_dl1})--(\ref{eqn_ns_dl2}), and (\ref{eqn_energy_dl}).

\subsection{Numerical implementation}
The numerical solution of the governing equations (\ref{eqn_continuity_dl1})--(\ref{eqn_ns_dl2}) and (\ref{eqn_energy_dl}) is obtained using COMSOL Multiphysics 5.4. 
The analysis domain is discretized using tetrahedral finite elements.
The scalar function $\psi (\mathbf{x})$ is transformed into a discrete form $\psi_i$ $(i=1, \ldots , n)$, where $n$ is the total number of nodal points in the design domain $D$. 
The initial values of $\psi_i$ are set to 0.5 uniformly. 
For the sensitivity analysis, we employ the adjoint method based on the discrete formulation described in the literature \citep{bendsoe2003topology}.

The optimization procedure in this study is described as follows:
\begin{description}
	\item[\textit{Step 1.}] The initial values of $\psi_i$ are set in $D$.
	\item[\textit{Step 2.}] The smoothed design variables $\gamma_i$ are derived using the Helmholtz-type filter in (\ref{eqn_filter}).
	\item[\textit{Step 3.}] The objective function in (\ref{eqn_optimization_problem}) is calculated by solving the governing equations in (\ref{eqn_continuity_dl1})--(\ref{eqn_ns_dl2}), and (\ref{eqn_energy_dl}). 
	\item[\textit{Step 4.}] If the value of $J$ is converged, the optimization process ends. Otherwise, the sensitivity is calculated using the adjoint method. 
	\item[\textit{Step 5.}] The design variables are updated using sequential linear programming (SLP), and the optimization procedure then returns to \textit{Step 2.}
\end{description}

\section{Numerical examples\label{sec_numerical}}
\subsection{Optimized flow fields\label{ssec_opt}}
Figure~\ref{fig4} shows the three-dimensional analysis model and its dimensions. The dimensions $L_x$, $L_y$, and $Lz$ are set to $4L$, $2L$, and $4L$, respectively. 
A symmetrical boundary condition is imposed in the $ZX$ plane.
The analysis domain is discretized using $5.3\times10^5$ tetrahedral elements.
The inlets and outlets of the two fluids are set as shown in Fig.~\ref{fig5}, which is referred to as counter flow. The dimensionless quantities are set to $Re_\mathrm{1}=100$, $Re_\mathrm{2}=100$, $Pr_\mathrm{f1}=7$, $Pr_\mathrm{f2}=7$, $Pr_\mathrm{s}=3.5$. The parameters regarding the interpolation functions are $\alpha_\mathrm{max}=1\times 10^4$, $q=0.01$, $s=0.1$. The filtering radius $R$ is set to $L/12$.

\begin{figure}[tbp]
\centering
\includegraphics[width=80mm]{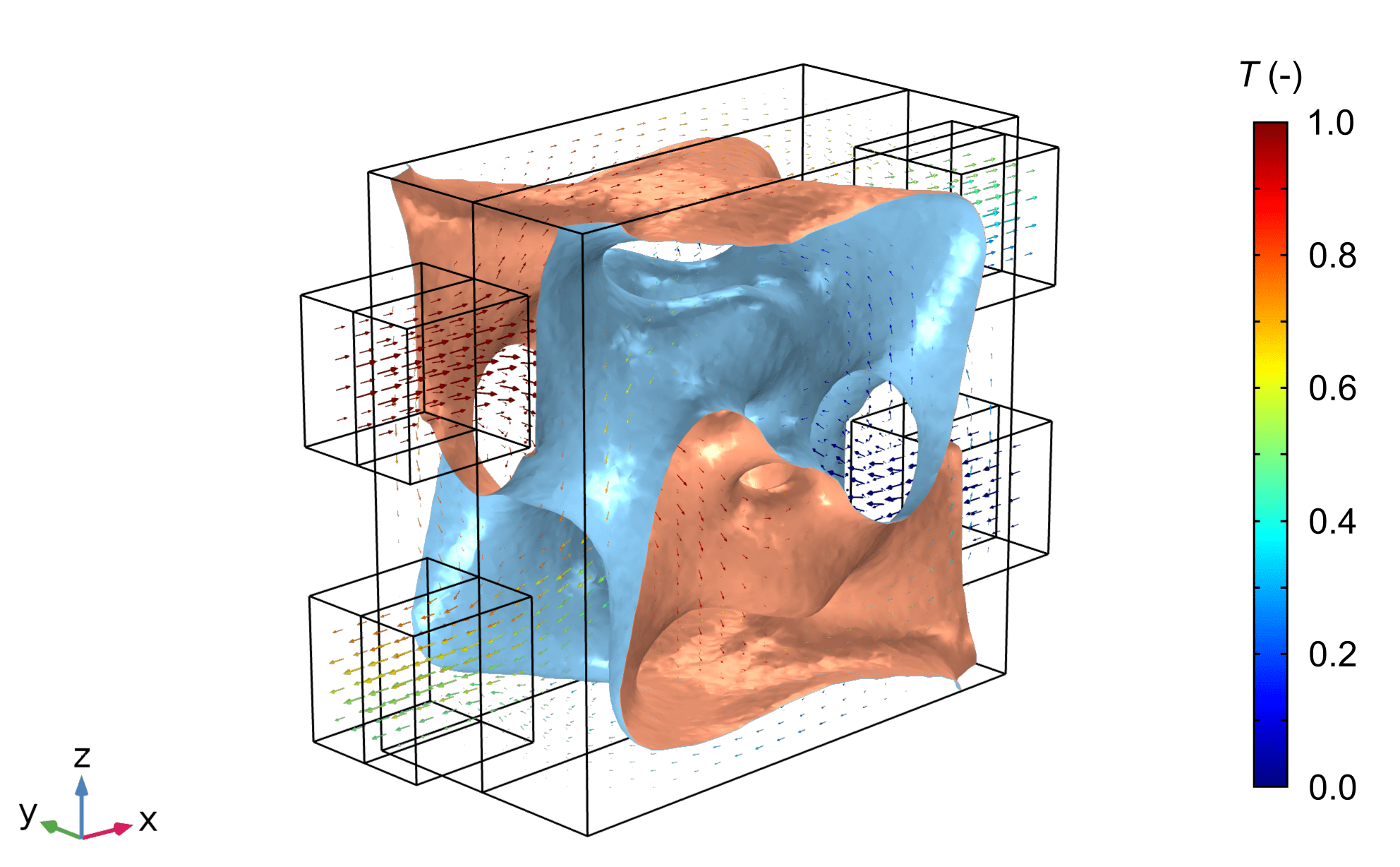}
\caption{Optimized result: isosurface of $\gamma=0.5$ and velocity vectors\label{fig6}}
\end{figure}

Figure~\ref{fig6} demonstrates the optimized result in which the isosurface of $\gamma=0.5$ and the velocity vectors are depicted. The separating wall is colored with orange on fluid 1 side, and light blue on fluid 2 side. The color of the arrows represents the temperature of the fluids. The optimized result of the half sectional view is shown in 
Fig.~\ref{fig7}. 
A validation of the optimized result is presented in the Appendix, where the optimization model is compared with the high-fidelity model that uses a body-fitted mesh.

The optimization history of $\gamma$ is illustrated in Fig.~\ref{fig8}, and the convergence history of $J$ is shown in Fig.~\ref{fig9}. From Fig.~\ref{fig8}, fluid 1 and fluid 2 domains appear as the optimization proceeds. Since the intermediate value of $\gamma$ exists between fluid 1 and fluid 2, the separating wall can be represented. The objective function value converges by approximately the 100th step, as depicted in Fig.~\ref{fig9}. 

Geometrically complex flow channels are obtained, as shown in Figs.~\ref{fig6} and \ref{fig7}. The flow channels are gradually curved and branched to circulate the fluids in the whole heat exchange domain, and fluid 1 and fluid 2 are placed alternately. This is because increasing the heat transfer area without the sharp bend is effective for improving the total heat transfer with low pressure loss. In addition, it is found that the outlet temperature of fluid 1 is lower than that of fluid 2. As illustrated in Fig.~\ref{fig7}, the downstream region of fluid 1 adjoins the upstream region of fluid 2 and vice versa. Thus, the fluids can exchange heat in the downstream region, whereas the heat is already transferred in the upstream branched region. This feature is unique to the counter flow and can enhance the heat transfer efficiency. 

\begin{figure}[tbp]
	\centering
	\includegraphics[width=80mm]{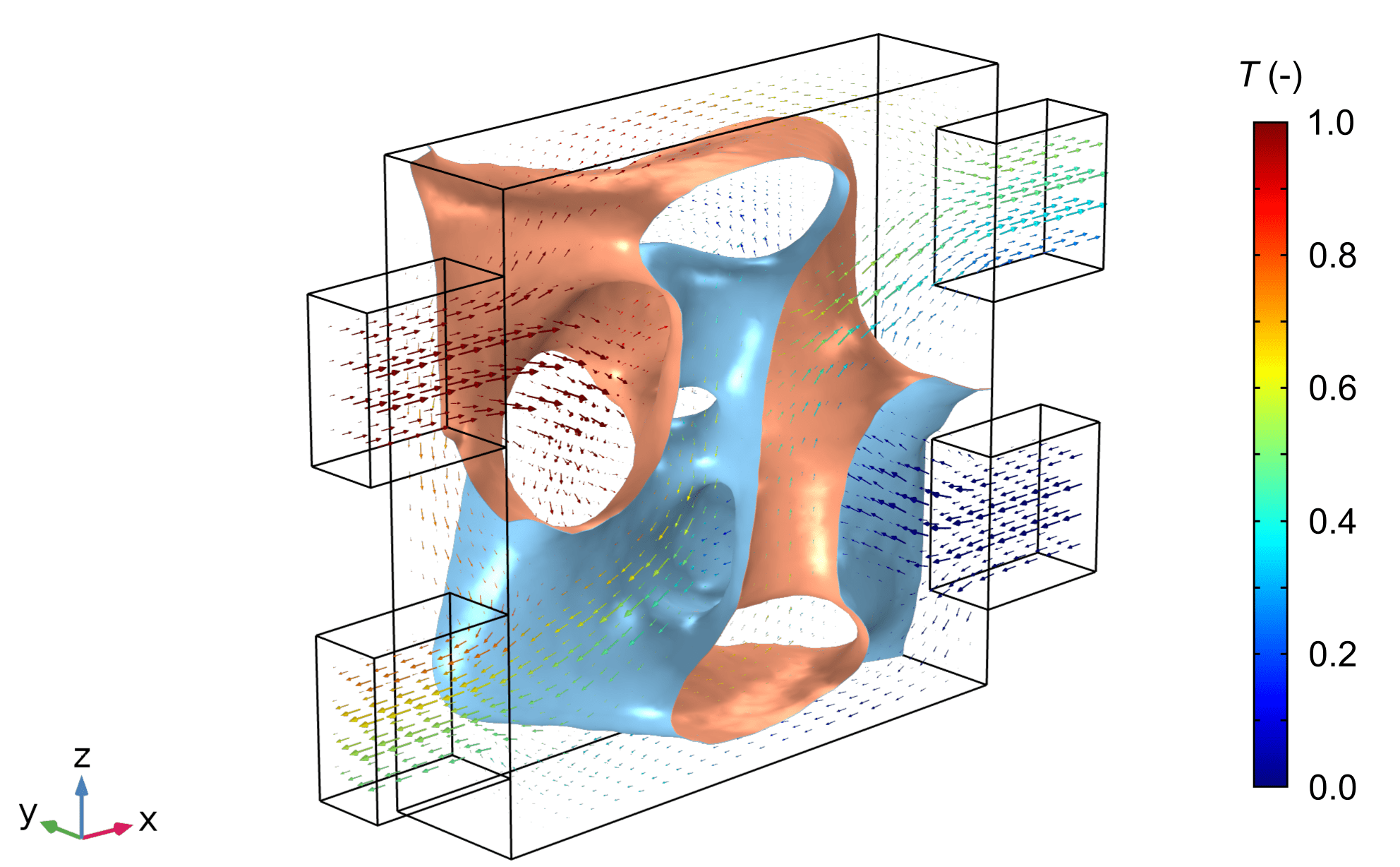}
	\caption{Half sectional view of optimized result in Fig.~\ref{fig6}\label{fig7}}
\end{figure}
\begin{figure*}[tbp]
	\centering
	\includegraphics[width=168mm]{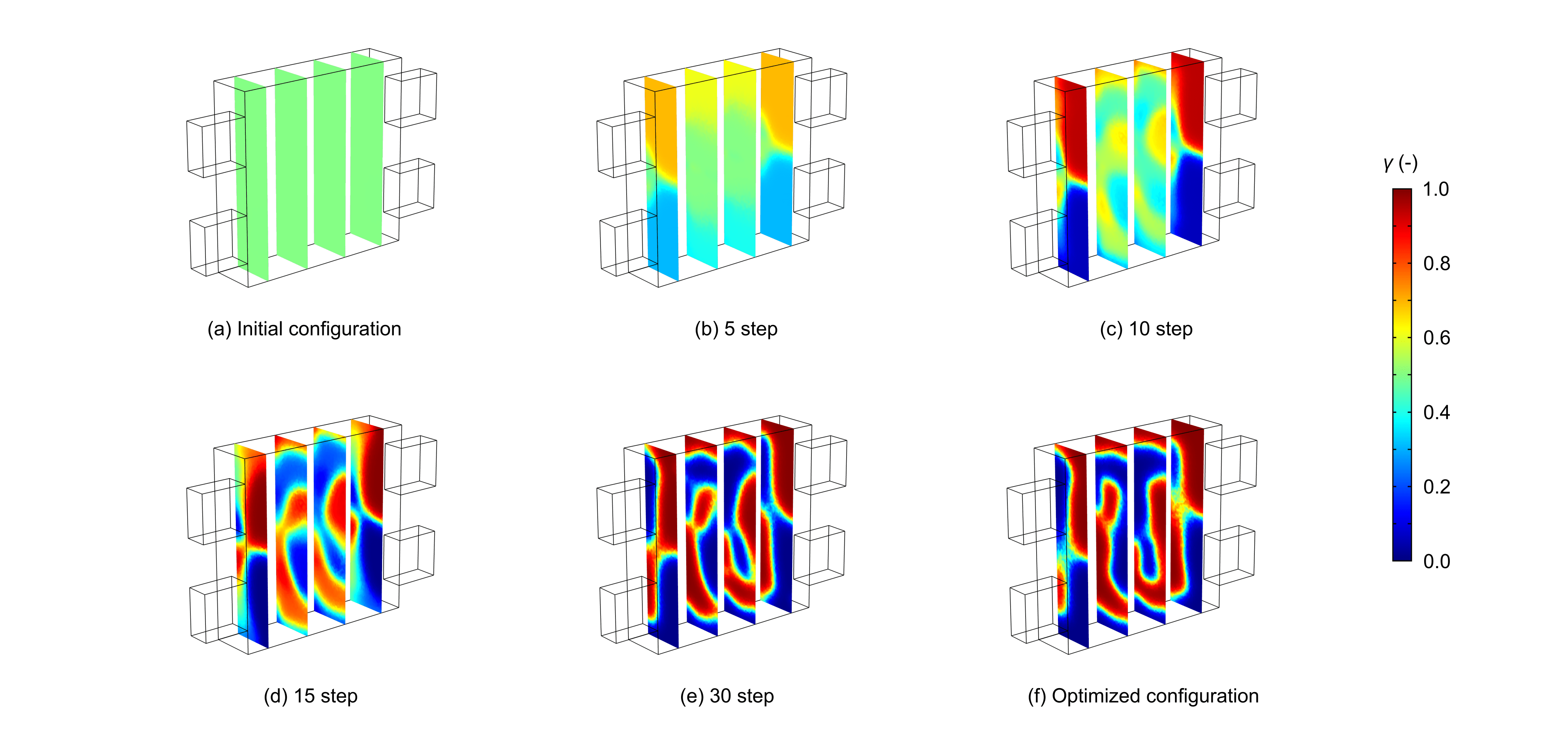}
	\caption{Optimization history of $\gamma$ in $XY$ slices\label{fig8}}
\end{figure*}
\begin{figure}[tbp]
	\centering
	\includegraphics[width=80mm]{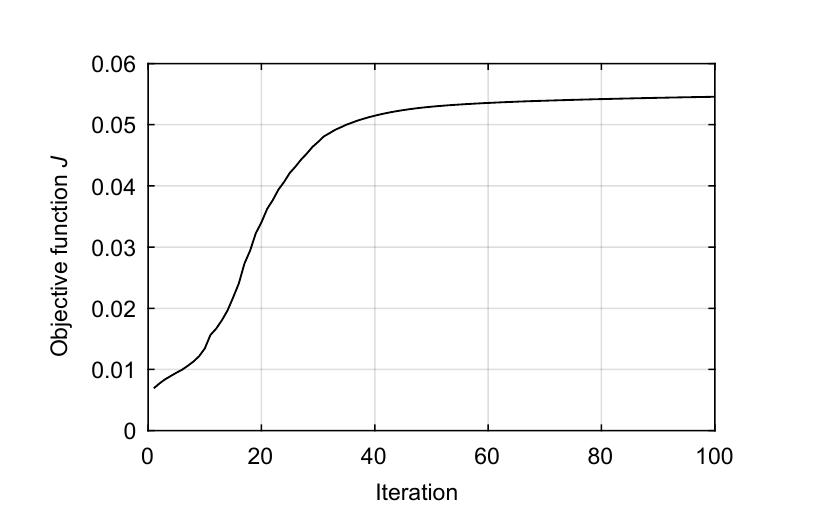}
	\caption{Convergence history of $J$\label{fig9}}
\end{figure}

\begin{table}[tbp]
	\centering
	\caption{Thermal-fluid performance of the optimized design and reference design\label{table3}}
	\begin{tabular}{c|ccc}
		\toprule
		& Total heat transfer & Flow rate of fluid 1 & Outlet temperature of fluid 1 \\
		& $J$ & $\dot{V}_1$ & $\bar{T}_\mathrm{out, 1}$ \\
		\midrule
		\midrule
		Optimized & 0.0538  & 0.0440  & 0.389  \\
		Reference & 0.0132 & 0.634 & 0.990 \\
		\bottomrule
	\end{tabular}%
\end{table}%

\subsection{Performance comparison with reference design}
\begin{figure}[tbp]
	\centering
	\includegraphics[width=80mm]{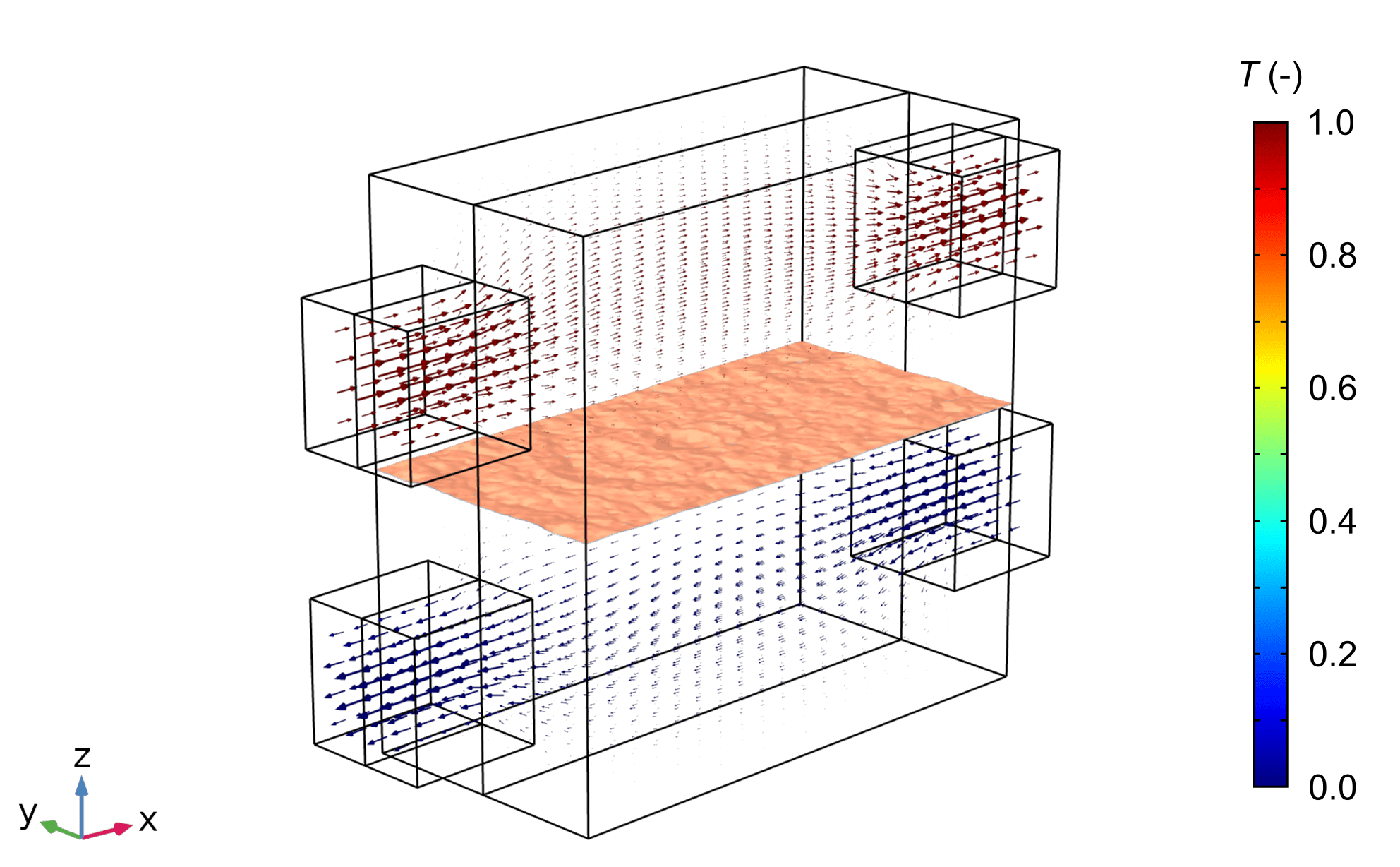}
	\caption{Reference design and its velocity vectors\label{fig10}}
\end{figure}

We now verify the performance of the optimized design shown in Fig.~\ref{fig6}, by comparison with the simple reference design that is composed of straight flow channels. Figure~\ref{fig10} illustrates the reference design and its velocity vector. It should be noted that the scale of the vector is not the same as that of the optimized design. 

Table~\ref{table3} shows the comparison of the total heat transfer $J$, the flow rate of fluid 1, i.e., $\dot{V}_1$, and the mean outlet temperature of fluid 1, i.e., $\bar{T}_\mathrm{out, 1}$. 
The flow rate of the reference design is over ten times higher than that of the optimized design. 
However, the mean outlet temperature is approximately the same as the inlet temperature in the reference design. 
Consequently, the mean outlet temperature is low in the optimized design. 
Thus, the total heat transfer is over four times higher than that of the reference design. 
Therefore, it is observed that the optimized design enhances heat transfer under the investigated operating condition. 

\subsection{Effects of design conditions}
\subsubsection{Effect of Reynolds number}
\begin{figure}[tbp]
	\centering
	\includegraphics[width=80mm]{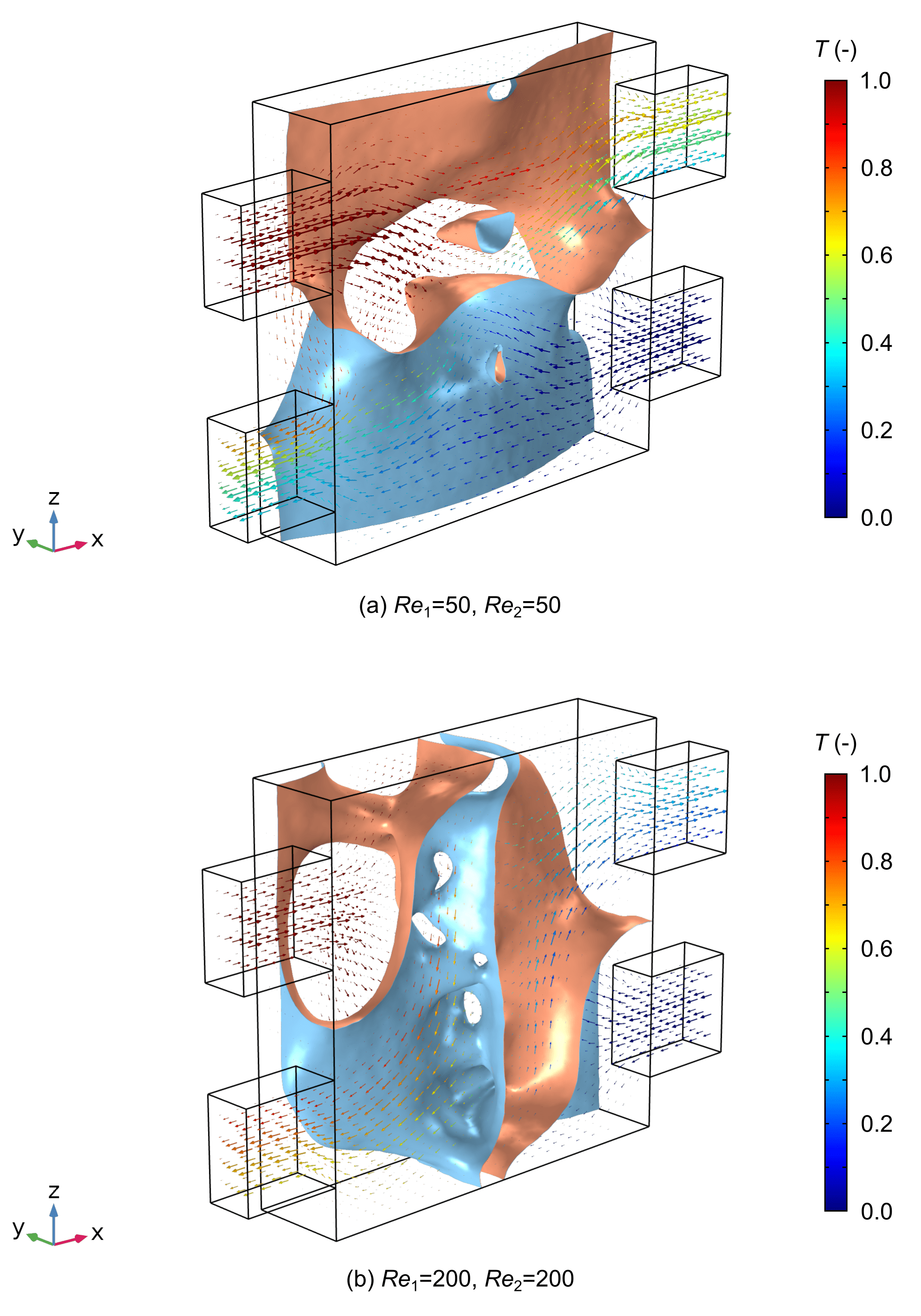}
	\caption{Optimized result for different Reynolds number in half sectional view: (a) $Re_1=50$, $Re_2=50$, (b) $Re_1=200$, $Re_2=200$\label{fig11}}
\end{figure}

Here, we investigate the effect of Reynolds number, $Re$, settings on the optimized configuration. In this study, the Reynolds number represents the pressure loss, because the characteristic speed is based on the pressure loss as defined in (\ref{eqn_U1}). Figure~\ref{fig11} illustrates the optimized results in the case of $Re_1=50$, $Re_2=50$ and the case of $Re_1=200$, $Re_2=200$. The parameters except for the Reynolds number are identical to the case of $Re_1=100$, $Re_2=100$ presented in Section~\ref{ssec_opt}. 

From Figs.~\ref{fig7} and \ref{fig11}, the number of branches increases as the Reynolds number increases. In Fig.~\ref{fig11}, a relatively simple structure is obtained in low $Re$. The heat transfer surface is approximately planar instead of the circular channels. Conversely, narrow branched channels are obtained for high $Re$. This is because the heat transfer area can be increased when the high pressure loss is allowed. In Fig.~\ref{fig11}(b), it can be observed that the outlet temperature of fluid 1 is significantly low for high $Re$, owing to the complex flow channels. These results indicate that the ideal topology of the two fluids varies by the Reynolds number setting.
\begin{figure}[tbp]
	\centering
	\includegraphics[width=80mm]{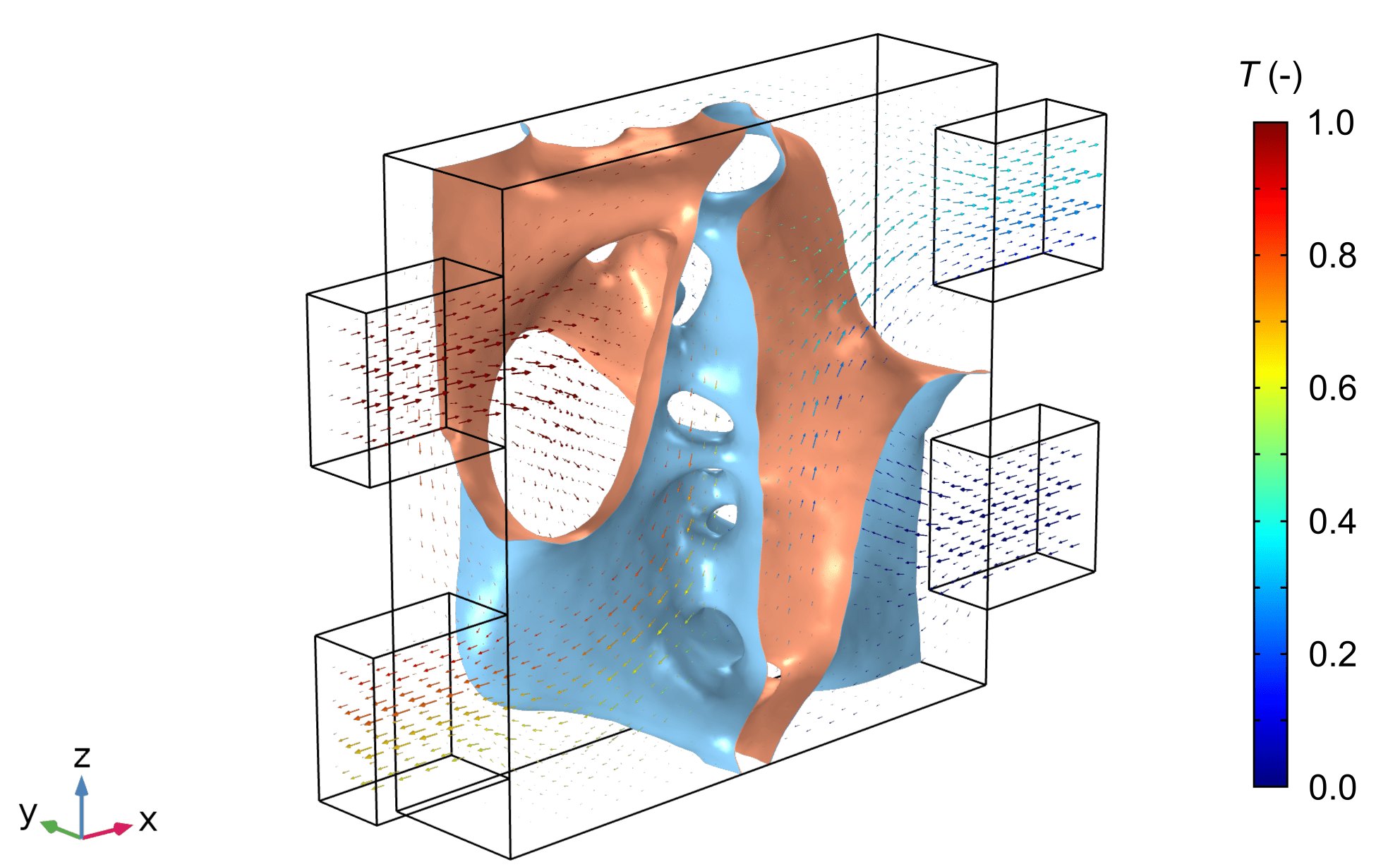}
	\caption{Optimized result for Prandtl number $Pr_\mathrm{f1}=14$, $Pr_\mathrm{f2}=14$, in half sectional view\label{fig12}}
\end{figure}
\begin{figure}[tbp]
	\centering
	\includegraphics[width=80mm]{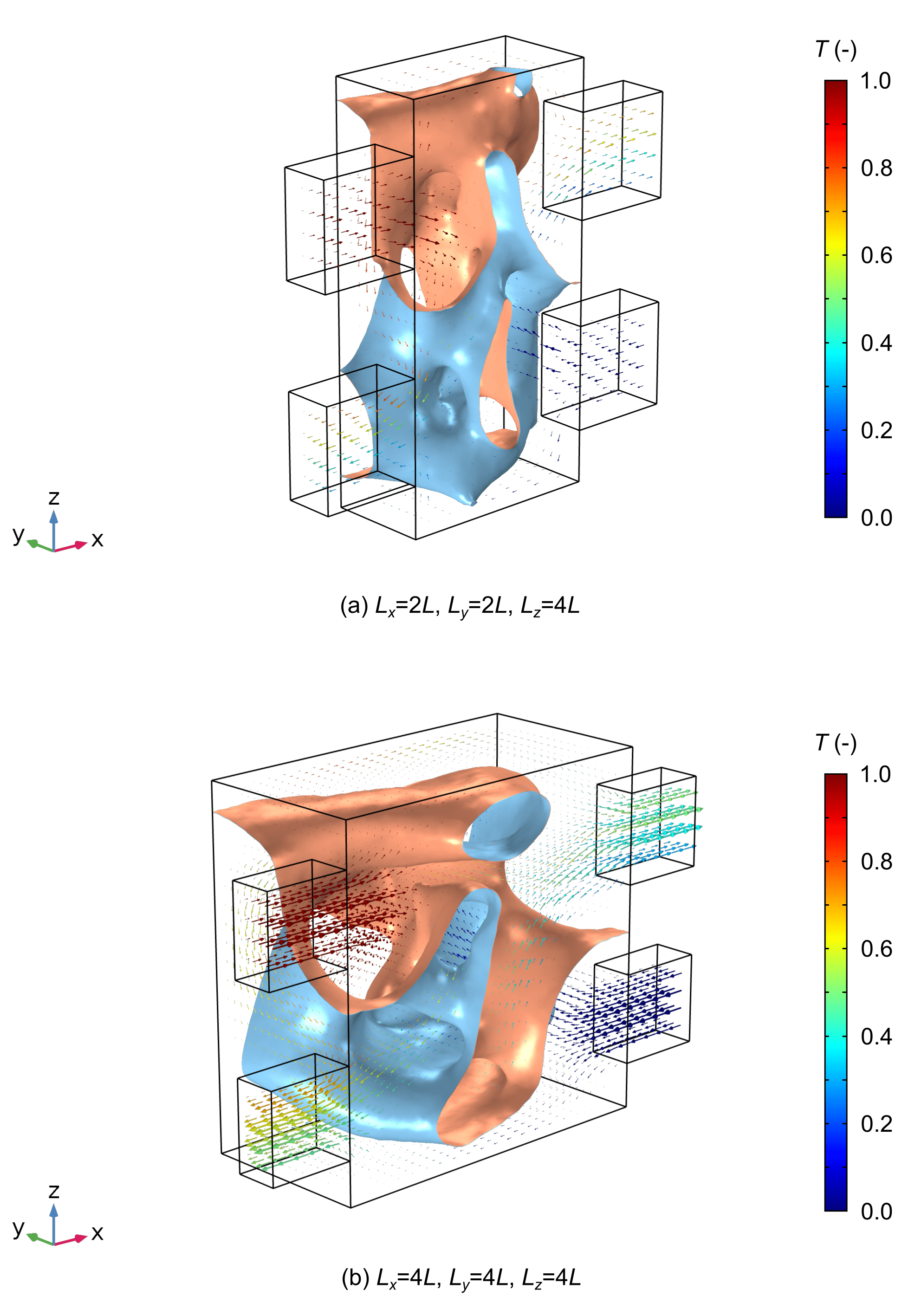}
	\caption{Optimized result for different design domain size in half sectional view: (a) $L_x=2L$, $L_y=2L$, $L_z=4L$, (b) $L_x=4L$, $L_y=4L$, $L_z=4L$\label{fig13}} 
\end{figure}
\begin{figure*}[tbp]
	\centering
	\includegraphics[width=160mm]{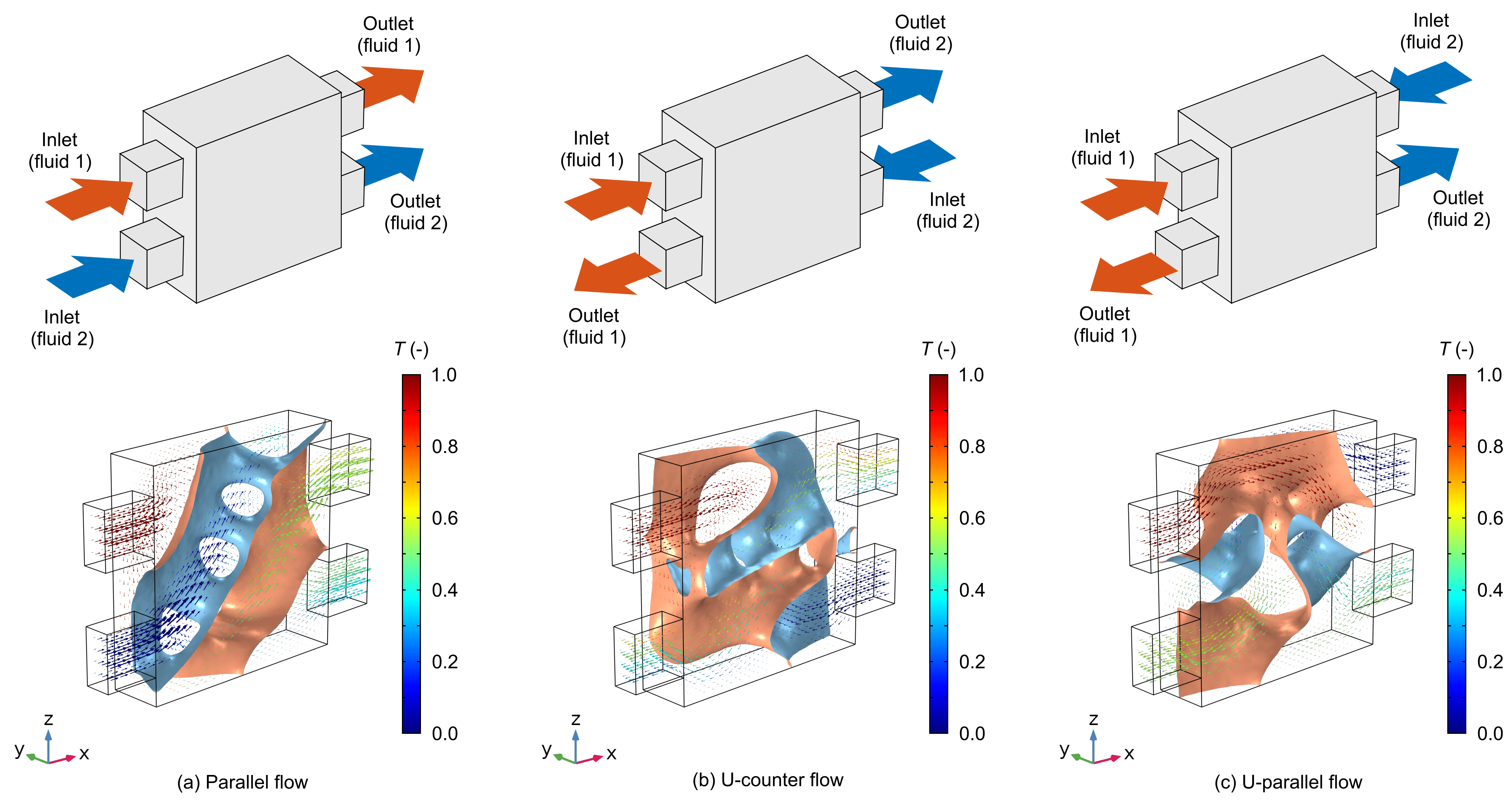}
	\caption{Schematic diagram and optimized result of several flow arrangements: (a) parallel flow, (b) U-counter flow, and (c) U-parallel flow\label{fig14}}
\end{figure*}

\subsubsection{Effect of Prandtl number}
Figure~\ref{fig12} shows the optimized results in a case of $Pr_\mathrm{f1}=14$, $Pr_\mathrm{f2}=14$. The parameters except for the Prandtl number are similar to the case of $Pr_\mathrm{f1}=7$, $Pr_\mathrm{f2}=7$ presented in Section~\ref{ssec_opt}. The Prandtl number shown in (\ref{eqn_dimensionless_temp}) is defined as the ratio of momentum diffusivity to thermal diffusivity. Hence, high Prandtl number results in low total heat transfer. 

In Fig.~\ref{fig12}, complex flow channels are obtained, which is similar to the case of $Re_1=200$, $Re_2=200$ shown in Fig.~\ref{fig11}(b). The reason for this similarity is related to the trade-off  between the total heat transfer and the pressure loss. Increasing the Reynolds number corresponds with the eased pressure loss limit. In contrast, increasing the Prandtl number corresponds with the strict requirement for total heat transfer. This can be explained by the fact that P\'{e}clet number of both cases in Fig.~\ref{fig11}(b) and Fig.~\ref{fig12} are identical; $Pe_\mathrm{f1}=1400$, $Pe_\mathrm{f2}=1400$.

\subsubsection{Effect of design domain size}
To investigate the effect of the design domain size, the optimization is performed in various $L_x$, $L_y$, and $L_z$. Figure~\ref{fig13} shows the optimized results for $L_x=2L$, $L_y=2L$, and $L_z=4L$, as well as for $L_x=4L$, $L_y=4L$, and $L_z=4L$. The parameters are similar to the case of $L_x=4L$, $L_y=2L$, and $L_z=4L$ presented in Section~\ref{ssec_opt}. 

Although the flow channels are placed alternately regardless of the design domain size, different features are obtained in each case. For Fig.~\ref{fig13}(a), the branched channels are in the $z$ direction instead of the $x$ direction as found in the cases of Figs.~\ref{fig7} and \ref{fig13}(b). Since the length in $x$ direction is short and the heat exchange domain is small, a relatively sharp bend is obtained. From Fig.~\ref{fig13}(b), thick and flat flow channels are realized when the length in $y$ direction is large. This is because the narrow flow channels are not necessary in case of a large heat exchange domain.

\subsubsection{Effect of flow arrangement}
Finally, the effect of the flow arrangement, i.e., the arrangement of the inlet and the outlet in each fluid, is investigated. Figure~\ref{fig14} displays the schematic diagram and the optimized result. Therefore, we denote the arrangement of (b) and (c) in Fig.~\ref{fig14} by ``U-counter'' and ``U-parallel'', respectively. 

From Fig.~\ref{fig14}(a), the flow channels of each fluid are placed alternately at right angles in parallel flow. It is discovered that the heat is mainly exchanged in the upstream region, and the temperature is nearly unchanged in the downstream region. This is because both fluids have approximately the same temperature after passing the branched section, and downstream fluid 1 cannot be adjacent to the upstream fluid 2, and vice versa, unlike in the case of counter flow. Similar features are found in Fig.~\ref{fig14}(c), U-parallel flow. In Fig.~\ref{fig14}(b), the high temperature difference can be maintained in the whole heat exchange surface, which is similar to counter flow. Based on these results, it is revealed that the outlet of the fluid 1 should be close to the inlet of the fluid 2, and vice versa, for efficient heat exchange. 

\section{Conclusions\label{sec_conclusion}}
In this paper, we proposed a density-based topology optimization method for two-fluid heat exchange problems.
The novel aspect of the paper is that a representation model involving three states (two types of fluids and a solid wall) is defined using a single design variable field without any penalty schemes. 
To this end, we introduced two types of fictitious force terms in each flow field so that the two fluids cannot essentially mix.

We formulated the optimization problem as a maximization problem of the heat transfer rate under the fixed pressure loss.
Based on the formulation, we investigated the effectiveness of the proposed approach through three-dimensional numerical examples.

In the numerical examples, we first showed the performance of the optimized design in comparison with a simple reference design composed of straight flow channels. 
It was found that the optimized design achieves a heat transfer performance over four times higher than that of the reference design.
We demonstrated the dependency of the optimized design with respect to the design conditions, i.e., Reynolds number, Prandtl number, design domain size, and flow arrangements.
Accordingly, it was found that the proposed approach can generate effective optimized designs for optimization problem concerning the two-fluid heat exchange system, at least under the operating conditions investigated in this paper. 

\section{ Replication of results}
The necessary information for replication of the results is presented in this paper. 
The interested reader may contact the corresponding author for further implementation details.

\section*{Acknowledgements}
This work was partially supported by JSPS KAKENHI Grant Number 18K13674.

\section*{Appendix: Validation via high-fidelity analysis}
Here, we conduct a validation of the optimization model via a high-fidelity analysis to demonstrate the possibility of the proposed method. 

To elucidate the physical phenomena more accurately, a projection technique is applied in the optimization model. This is because the smoothed design variable field $\gamma$ is not strictly 1 and 0 in fluid 1 and 2 domains, owing to the Helmholtz-type filter in (\ref{eqn_filter}). 
Therefore, the projection function is introduced as follows \citep{wang2011projection}:
\begin{equation}
\hat{\gamma} = \frac{\tanh (\beta \eta) + \tanh (\beta(\gamma-\eta))}{\tanh (\beta \eta) + \tanh (\beta (1-\eta))},
\end{equation}
\cl{black}
where $\hat{\gamma}$ is the material distribution after the projection, and $\beta$ and $\eta$ are the tuning parameters that control the function shape. In this example, $\beta$ and $\eta$ are set to 8 and 0.5, respectively. 

We use the same design setting as in Section~\ref{ssec_opt}. The projection function is applied from 101 steps to 200 steps, i.e., the additional optimization process is performed for 100 steps, using the result in Fig.~\ref{fig7} as the initial configuration.

\begin{figure}[tbp]
	\centering
	\includegraphics[width=80mm]{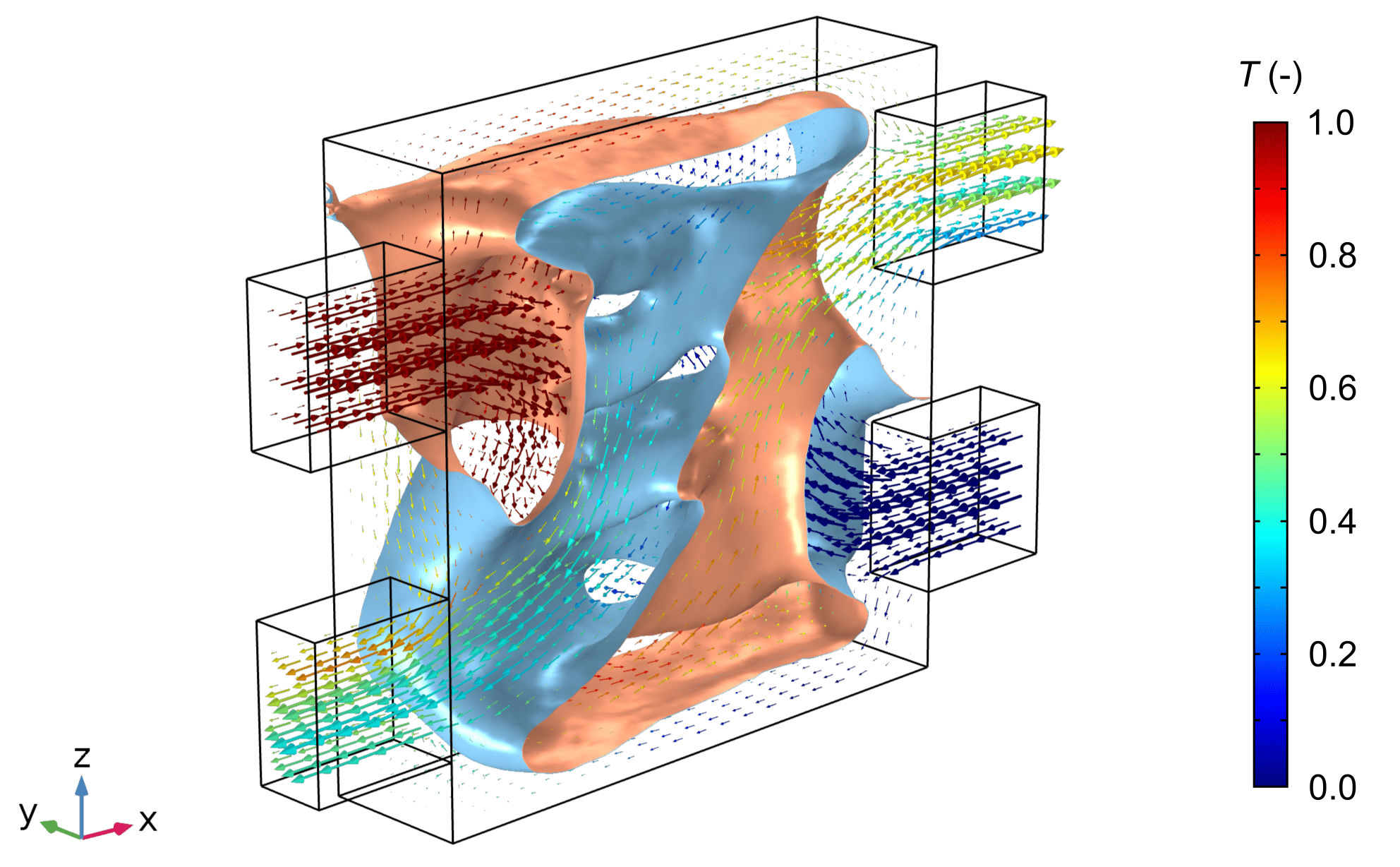}
	\caption{Optimized result using projection scheme in half sectional view\label{fig15}}
\end{figure}

Figure~\ref{fig15} shows the optimized result. It should be noted that the scale of the vector is not the same as Fig.~\ref{fig7}. As indicated in Fig.~\ref{fig15}, the optimized configuration is similar to the case without projection although the flow channels become more narrow and complex. The fluids can flow with low drag forces because the inverse permeability in the fluid domains is reduced by the projection. Thus, a more complex configuration is allowed. 
\begin{figure}[tbp]
	\centering
	\includegraphics[width=80mm]{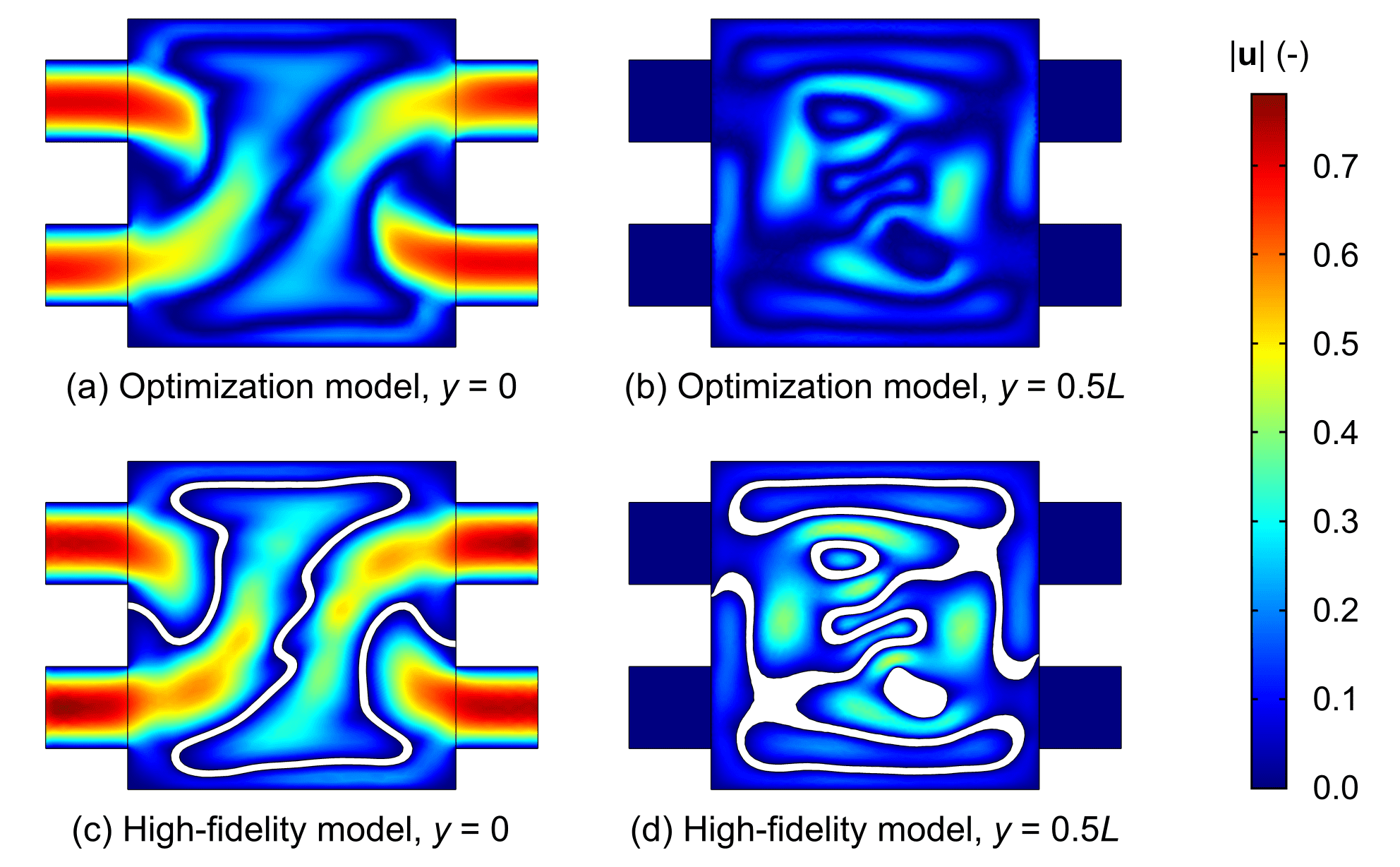}
	\caption{Velocity magnitude in $ZX$ plane: (a) Optimization model, $y=0$, (b) Optimization model, $y=0.5L$, (c) High-fidelity model, $y=0$, (d) High-fidelity model, $y=0.5L$\label{fig16}}
\end{figure}
\begin{figure}[tbp]
	\centering
	\includegraphics[width=80mm]{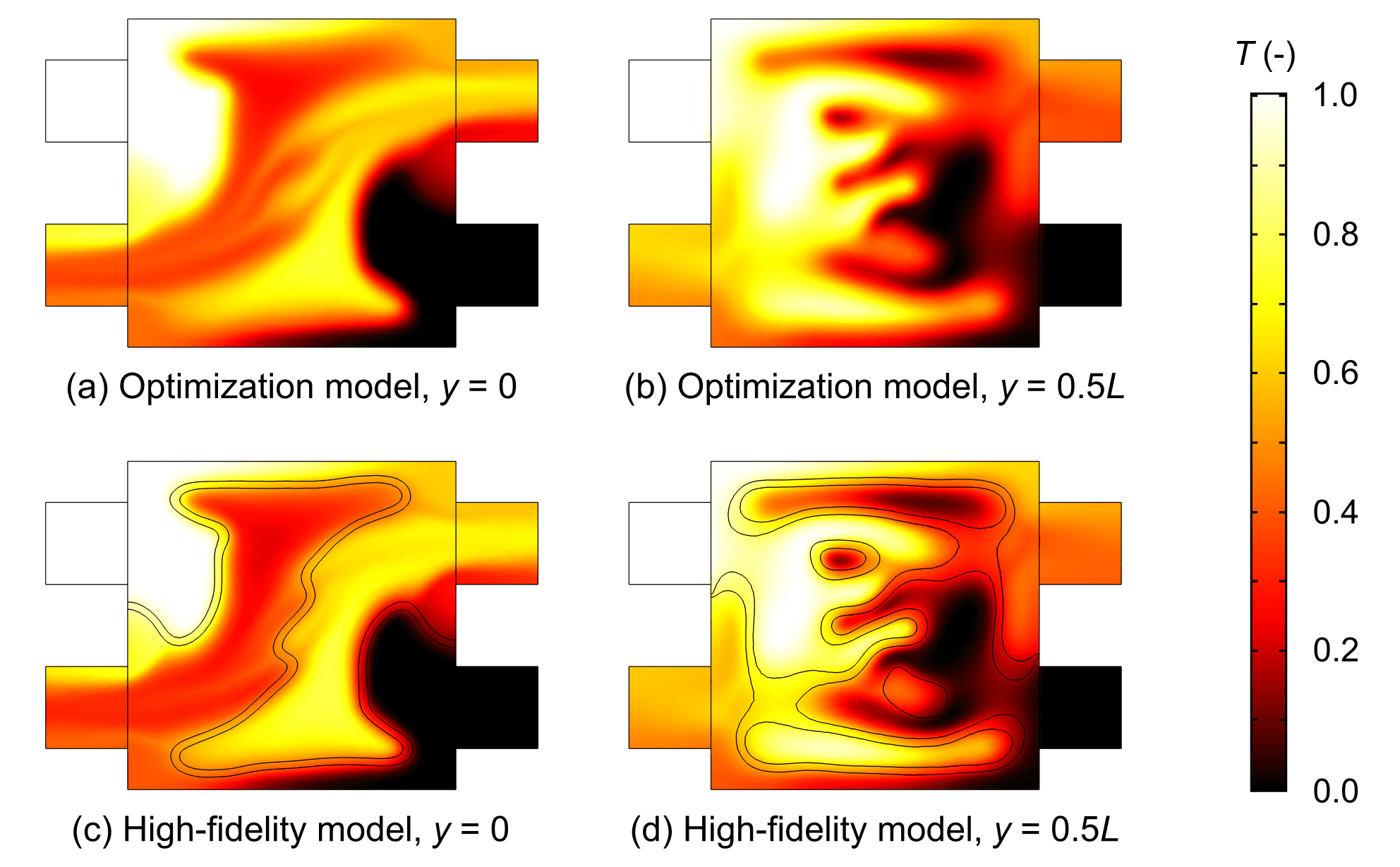}
	\caption{Temperature in $ZX$ plane: (a) Optimization model, $y=0$, (b) Optimization model, $y=0.5L$, (c) High-fidelity model, $y=0$, (d) High-fidelity model, $y=0.5L$\label{fig17}}
\end{figure}

In the high-fidelity analysis, the boundaries are extracted from the isosurface of the material distribution to use a body-fitted mesh.
 In this example, the wall thickness is set to approximately $0.08L$. To elucidate the velocity and temperature gradients, finer prism cells are used in the near-wall regions. Other settings for analysis are similar to those in the optimization model. 

Figures~\ref{fig16} and \ref{fig17} show the velocity magnitude and the temperature of the optimization and the high-fidelity models. Thus, the agreement between the optimization and the high-fidelity models is demonstrated. 
Table \ref{tab:addlabel} shows the comparison results in terms of the total heat transfer, $J$, the flow rate of fluid 1, $\dot{V}_1$, and the mean outlet temperature of fluid 1, $\bar{T}_\mathrm{out, 1}$.
In addition, the errors of $J$, $\dot{V}_1$, and $\bar{T}_\mathrm{out, 1}$, are 5.99\%, 2.82\%, and 7.99\%, respectively. 
These results reveal that the optimization model can predict the thermal-fluid characteristics.

It should be noted that although one example in Section~\ref{ssec_opt} is treated for the validation here, the projection technique as aforedescribed is necessary to decrease the discrepancy between the optimization model and the high-fidelity model.

\begin{table}[htbp]
	\centering
	\caption{Comparison of total heat transfer, the flow rate of fluid 1, and the mean outlet temperature of fluid 1}
	\begin{tabular}{c|ccc}
		\toprule
		& Total heat transfer & Flow rate of fluid 1 & Outlet temperature of fluid 1 \\
		& $J$   & $\dot{V}_1$ & $\bar{T}_\mathrm{out, 1}$ \\
		\midrule
		\midrule
		Optimization & 0.1699  & 0.1684  & 0.495  \\
		High-fidelity & 0.1603  & 0.1733 & 0.538  \\
		\bottomrule
	\end{tabular}%
	\label{tab:addlabel}%
\end{table}%

\section*{Conflict of interest}
The authors declare that they have no conflict of interest.

\bibliographystyle{spbasic}
\bibliography{reference}

\begin{thebibliography}{41}
\providecommand{\natexlab}[1]{#1}
\providecommand{\url}[1]{{#1}}
\providecommand{\urlprefix}{URL }
\expandafter\ifx\csname urlstyle\endcsname\relax
  \providecommand{\doi}[1]{DOI~\discretionary{}{}{}#1}\else
  \providecommand{\doi}{DOI~\discretionary{}{}{}\begingroup
  \urlstyle{rm}\Url}\fi
\providecommand{\eprint}[2][]{\url{#2}}

\bibitem[{Alexandersen and Andreasen(2020)}]{alexandersen2020review}
Alexandersen J, Andreasen CS (2020) A review of topology optimisation for
  fluid-based problems. Fluids 5(1):29

\bibitem[{Alexandersen et~al.(2014)Alexandersen, Aage, Andreasen, and
  Sigmund}]{alexandersen2014topology}
Alexandersen J, Aage N, Andreasen CS, Sigmund O (2014) Topology optimisation
  for natural convection problems. International Journal for Numerical Methods
  in Fluids 76(10):699--721

\bibitem[{Alexandersen et~al.(2016)Alexandersen, Sigmund, and
  Aage}]{alexandersen2016large}
Alexandersen J, Sigmund O, Aage N (2016) Large scale three-dimensional topology
  optimisation of heat sinks cooled by natural convection. International
  Journal of Heat and Mass Transfer 100:876--891

\bibitem[{Andreasen et~al.(2009)Andreasen, Gersborg, and
  Sigmund}]{andreasen2009topology}
Andreasen CS, Gersborg AR, Sigmund O (2009) Topology optimization of
  microfluidic mixers. International Journal for Numerical Methods in Fluids
  61(5):498--513

\bibitem[{Asmussen et~al.(2019)Asmussen, Alexandersen, Sigmund, and
  Andreasen}]{asmussen2019poor}
Asmussen J, Alexandersen J, Sigmund O, Andreasen CS (2019) A ``poor man's''
  approach to topology optimization of natural convection problems. Structural
  and Multidisciplinary Optimization 59(4):1105--1124

\bibitem[{Bends{\o}e and Kikuchi(1988)}]{bendsoe1988generating}
Bends{\o}e MP, Kikuchi N (1988) Generating optimal topologies in structural
  design using a homogenization method. Computer Methods in Applied Mechanics
  and Engineering 71(2):197--224

\bibitem[{Bends{\o}e and Sigmund(1999)}]{bendsoe1999material}
Bends{\o}e MP, Sigmund O (1999) Material interpolation schemes in topology
  optimization. Archive of applied mechanics 69(9-10):635--654

\bibitem[{Bends{\o}e and Sigmund(2003)}]{bendsoe2003topology}
Bends{\o}e MP, Sigmund O (2003) Topology optimization: theory, methods and
  applications. Springer

\bibitem[{Borrvall and Petersson(2003)}]{borrvall2003topology}
Borrvall T, Petersson J (2003) Topology optimization of fluids in {Stokes}
  flow. International Journal for Numerical Methods in Fluids 41(1):77--107

\bibitem[{Chen et~al.(2019)Chen, Yaji, Yamasaki, Tsushima, and
  Fujita}]{chen2019computational}
Chen CH, Yaji K, Yamasaki S, Tsushima S, Fujita K (2019) Computational design
  of flow fields for vanadium redox flow batteries via topology optimization.
  Journal of Energy Storage 26:100990

\bibitem[{Coffin and Maute(2016)}]{coffin2016level}
Coffin P, Maute K (2016) A level-set method for steady-state and transient
  natural convection problems. Structural and Multidisciplinary Optimization
  53(5):1047--1067

\bibitem[{Deaton and Grandhi(2014)}]{deaton2014survey}
Deaton JD, Grandhi RV (2014) A survey of structural and multidisciplinary
  continuum topology optimization: post 2000. Structural and Multidisciplinary
  Optimization 49(1):1--38

\bibitem[{Deng et~al.(2018)Deng, Zhou, Liu, Wu, Qian, and
  Korvink}]{deng2018topology}
Deng Y, Zhou T, Liu Z, Wu Y, Qian S, Korvink JG (2018) Topology optimization of
  electrode patterns for electroosmotic micromixer. International Journal of
  Heat and Mass Transfer 126:1299--1315

\bibitem[{Dilgen et~al.(2018)Dilgen, Dilgen, Fuhrman, Sigmund, and
  Lazarov}]{dilgen2018density}
Dilgen SB, Dilgen CB, Fuhrman DR, Sigmund O, Lazarov BS (2018) Density based
  topology optimization of turbulent flow heat transfer systems. Structural and
  Multidisciplinary Optimization 57(5):1905--1918

\bibitem[{Gersborg-Hansen et~al.(2005)Gersborg-Hansen, Sigmund, and
  Haber}]{gersborg2005topology}
Gersborg-Hansen A, Sigmund O, Haber RB (2005) Topology optimization of channel
  flow problems. Structural and Multidisciplinary Optimization 30(3):181--192

\bibitem[{Guo et~al.(2018)Guo, Zhang, and Smith}]{guo2018design}
Guo K, Zhang N, Smith R (2018) Design optimisation of multi-stream plate fin
  heat exchangers with multiple fin types. Applied Thermal Engineering
  131:30--40

\bibitem[{Haertel and Nellis(2017)}]{haertel2017fully}
Haertel JH, Nellis GF (2017) {A fully developed flow thermofluid model for
  topology optimization of 3D-printed air-cooled heat exchangers}. Applied
  Thermal Engineering 119:10--24

\bibitem[{Hilbert et~al.(2006)Hilbert, Janiga, Baron, and
  Th{\'e}venin}]{hilbert2006multi}
Hilbert R, Janiga G, Baron R, Th{\'e}venin D (2006) Multi-objective shape
  optimization of a heat exchanger using parallel genetic algorithms.
  International Journal of Heat and Mass Transfer 49(15-16):2567--2577

\bibitem[{Kambampati and Kim(2020)}]{kambampati2020level}
Kambampati S, Kim HA (2020) {Level set topology optimization of cooling
  channels using the Darcy flow model}. Structural and Multidisciplinary
  Optimization 61:1345--1361

\bibitem[{Kanaris et~al.(2009)Kanaris, Mouza, and Paras}]{kanaris2009optimal}
Kanaris A, Mouza A, Paras S (2009) Optimal design of a plate heat exchanger
  with undulated surfaces. International Journal of Thermal Sciences
  48(6):1184--1195

\bibitem[{Kawamoto et~al.(2011)Kawamoto, Matsumori, Yamasaki, Nomura, Kondoh,
  and Nishiwaki}]{kawamoto2011heaviside}
Kawamoto A, Matsumori T, Yamasaki S, Nomura T, Kondoh T, Nishiwaki S (2011)
  Heaviside projection based topology optimization by a {PDE}-filtered scalar
  function. Structural and Multidisciplinary Optimization 44(1):19--24

\bibitem[{Kobayashi et~al.(2019)Kobayashi, Yaji, Yamasaki, and
  Fujita}]{kobayashi2019freeform}
Kobayashi H, Yaji K, Yamasaki S, Fujita K (2019) Freeform winglet design of
  fin-and-tube heat exchangers guided by topology optimization. Applied Thermal
  Engineering 161:114020

\bibitem[{Koga et~al.(2013)Koga, Lopes, Nova, De~Lima, and
  Silva}]{koga2013development}
Koga AA, Lopes ECC, Nova HFV, De~Lima CR, Silva ECN (2013) Development of heat
  sink device by using topology optimization. International Journal of Heat and
  Mass Transfer 64:759--772

\bibitem[{Lazarov and Sigmund(2011)}]{lazarov2011filters}
Lazarov BS, Sigmund O (2011) Filters in topology optimization based on
  {Helmholtz}-type differential equations. International Journal for Numerical
  Methods in Engineering 86(6):765--781

\bibitem[{Lin et~al.(2015)Lin, Zhao, Guest, Weihs, and Liu}]{lin2015topology}
Lin S, Zhao L, Guest JK, Weihs TP, Liu Z (2015) Topology optimization of
  fixed-geometry fluid diodes. Journal of Mechanical Design 137(8)

\bibitem[{Matsumori et~al.(2013)Matsumori, Kondoh, Kawamoto, and
  Nomura}]{matsumori2013topology}
Matsumori T, Kondoh T, Kawamoto A, Nomura T (2013) Topology optimization for
  fluid-thermal interaction problems under constant input power. Structural and
  Multidisciplinary Optimization 47(4):571--581

\bibitem[{Okkels and Bruus(2007)}]{okkels2007scaling}
Okkels F, Bruus H (2007) Scaling behavior of optimally structured catalytic
  microfluidic reactors. Physical Review E 75(1):016301

\bibitem[{Olesen et~al.(2006)Olesen, Okkels, and Bruus}]{olesen2006high}
Olesen LH, Okkels F, Bruus H (2006) A high-level programming-language
  implementation of topology optimization applied to steady-state
  {Navier}--{Stokes} flow. International Journal for Numerical Methods in
  Engineering 65(7):975--1001

\bibitem[{Pollini et~al.(2020)Pollini, Sigmund, Andreasen, and
  Alexandersen}]{pollini2020poor}
Pollini N, Sigmund O, Andreasen CS, Alexandersen J (2020) A ``poor man's''
  approach for high-resolution three-dimensional topology design for natural
  convection problems. Advances in Engineering Software 140:102736

\bibitem[{Sato et~al.(2017)Sato, Yaji, Izui, Yamada, and
  Nishiwaki}]{sato2017topology}
Sato Y, Yaji K, Izui K, Yamada T, Nishiwaki S (2017) {Topology optimization of
  a no-moving-part valve incorporating Pareto frontier exploration}. Structural
  and Multidisciplinary Optimization 56(4):839--851

\bibitem[{Scheithauer et~al.(2018)Scheithauer, Schwarzer, Moritz, and
  Michaelis}]{scheithauer2018additive}
Scheithauer U, Schwarzer E, Moritz T, Michaelis A (2018) {Additive
  manufacturing of ceramic heat exchanger: opportunities and limits of the
  lithography-based ceramic manufacturing (LCM)}. Journal of Materials
  Engineering and Performance 27(1):14--20

\bibitem[{Shah and Sekulic(2003)}]{shah2003fundamentals}
Shah RK, Sekulic DP (2003) Fundamentals of heat exchanger design. John Wiley \&
  Sons

\bibitem[{Sigmund and Maute(2013)}]{sigmund2013topology}
Sigmund O, Maute K (2013) Topology optimization approaches. Structural and
  Multidisciplinary Optimization 48(6):1031--1055

\bibitem[{de~Souza and Silva(2020)}]{detopology}
de~Souza EM, Silva ECN (2020) Topology optimization applied to the design of
  actuators driven by pressure loads. Structural and Multidisciplinary
  Optimization

\bibitem[{Tawk et~al.(2019)Tawk, Ghannam, and Nemer}]{tawk2019topology}
Tawk R, Ghannam B, Nemer M (2019) Topology optimization of heat and mass
  transfer problems in two fluids---one solid domains. Numerical Heat Transfer,
  Part B: Fundamentals 76(3):130--151

\bibitem[{Wang et~al.(2011)Wang, Lazarov, and Sigmund}]{wang2011projection}
Wang F, Lazarov BS, Sigmund O (2011) On projection methods, convergence and
  robust formulations in topology optimization. Structural and
  Multidisciplinary Optimization 43(6):767--784

\bibitem[{Yaji et~al.(2015)Yaji, Yamada, Kubo, Izui, and
  Nishiwaki}]{yaji2015topology}
Yaji K, Yamada T, Kubo S, Izui K, Nishiwaki S (2015) A topology optimization
  method for a coupled thermal--fluid problem using level set boundary
  expressions. International Journal of Heat and Mass Transfer 81:878--888

\bibitem[{Yaji et~al.(2018{\natexlab{a}})Yaji, Ogino, Chen, and
  Fujita}]{yaji2018large}
Yaji K, Ogino M, Chen C, Fujita K (2018{\natexlab{a}}) Large-scale topology
  optimization incorporating local-in-time adjoint-based method for unsteady
  thermal-fluid problem. Structural and Multidisciplinary Optimization
  58(2):817--822

\bibitem[{Yaji et~al.(2018{\natexlab{b}})Yaji, Yamasaki, Tsushima, Suzuki, and
  Fujita}]{yaji2018topology}
Yaji K, Yamasaki S, Tsushima S, Suzuki T, Fujita K (2018{\natexlab{b}})
  Topology optimization for the design of flow fields in a redox flow battery.
  Structural and Multidisciplinary Optimization 57(2):535--546

\bibitem[{Yaji et~al.(2020)Yaji, Yamasaki, and Fujita}]{yaji2020multifidelity}
Yaji K, Yamasaki S, Fujita K (2020) Multifidelity design guided by topology
  optimization. Structural and Multidisciplinary Optimization 61(3):1071--1085

\bibitem[{Zhao et~al.(2018)Zhao, Zhou, Sigmund, and Andreasen}]{zhao2018poor}
Zhao X, Zhou M, Sigmund O, Andreasen CS (2018) {A ``poor man's approach'' to
  topology optimization of cooling channels based on a Darcy flow model}.
  International Journal of Heat and Mass Transfer 116:1108--1123

\end{thebibliography}

\end{document}